\documentclass[letterpaper]{article} 
\usepackage{aaai25}  
\usepackage{times}  
\usepackage{helvet}  
\usepackage{courier}  
\usepackage[hyphens]{url}  
\usepackage{graphicx} 
\usepackage{natbib}  
\usepackage{caption} 
\usepackage{algorithm}
\usepackage{algorithmic}

\usepackage{subfigure}
\usepackage{xcolor}
\usepackage{tabularray}

\usepackage{amsmath}
\usepackage{amssymb}
\usepackage{booktabs}
\usepackage{multirow}
\newcommand{\omegaV}{\boldsymbol{\omega}}
\newcommand{\thetaV}{\boldsymbol{\theta}}
\newcommand{\ThetaV}{\boldsymbol{\Theta}}
\newcommand{\oV}{\boldsymbol{o}}
\newcommand{\OV}{\boldsymbol{O}}
\newcommand{\mysubsubsection}[1]{\noindent \textbf{#1}}

\newtheorem{definition}{Definition}

\newcommand{\ignore}[1]{}
\usepackage{newfloat}
\usepackage{listings}
\usepackage{subcaption}
\DeclareCaptionStyle{ruled}{labelfont=normalfont,labelsep=colon,strut=off} 
\floatstyle{ruled}
\newfloat{listing}{tb}{lst}{}
\floatname{listing}{Listing}
\title{Online Guidance Graph Optimization for Lifelong Multi-Agent Path Finding}
\author{
    Hongzhi Zang\textsuperscript{\rm 1}\equalcontrib, Yulun Zhang\textsuperscript{\rm 2}\equalcontrib, He Jiang\textsuperscript{\rm 2}, Zhe Chen\textsuperscript{\rm 3}, \\
    Daniel Harabor\textsuperscript{\rm 3}, Peter J. Stuckey\textsuperscript{\rm 3}, Jiaoyang Li\textsuperscript{\rm 2}\
}
\affiliations{
    \textsuperscript{\rm 1} Institute for Interdisciplinary Information Sciences, Tsinghua University, Beijing, 100084, China\\
    zanghz21@mails.tsinghua.edu.cn\\
    \textsuperscript{\rm 2} Robotics Institute, Carnegie Mellon University, Pittsburgh, PA15213, USA \\
    \{yulunzhang, hejiangrivers, jiaoyangli\}@cmu.edu\\
    \textsuperscript{\rm 3} Department of Data Science and AI, Monash University, Melbourne, 3800 VIC, Australia \\
    \{zhe.chen, daniel.harabor, peter.stuckey\}@monash.edu
}

\usepackage{cleveref} 
\Crefname{ALC@unique}{Line}{Line}

\begin{document}

\maketitle

\begin{abstract}
We study the problem of optimizing a guidance policy capable of dynamically guiding the agents for lifelong Multi-Agent Path Finding based on real-time traffic patterns. Multi-Agent Path Finding (MAPF) focuses on moving multiple agents from their starts to goals without collisions. Its lifelong variant, LMAPF, continuously assigns new goals to agents. In this work, we focus on improving the solution quality of PIBT, a state-of-the-art rule-based LMAPF algorithm, by optimizing a policy to generate adaptive guidance. We design two pipelines to incorporate guidance in PIBT in two different ways. We demonstrate the superiority of the optimized policy over both static guidance and human-designed policies. Additionally, we explore scenarios where task distribution changes over time, a challenging yet common situation in real-world applications that is rarely explored in the literature.
\end{abstract}

%
\begin{links}
    \link{Code}{https://github.com/zanghz21/OnlineGGO}
    \link{arXiv}{https://arxiv.org/abs/2411.16506}
\end{links}


\section{Introduction}


We study the problem of optimizing a \textit{guidance policy} capable of dynamically updating guidance graphs with optimized edge weights to guide the agents' movement and improve throughput in lifelong Multi-Agent Path Finding.
Multi-Agent Path Finding (MAPF)~\cite{SternSoCS19} aims to compute collision-free paths for agents from their starts to goals. The lifelong variant of MAPF (LMAPF) constantly assigns new goals to agents upon arriving at their current goals and aims to maximize \emph{throughput}, namely the number of goals reached per timestep. LMAPF has wide applications in automated warehouses~\cite{Li2020LifelongMP,zhangLayout23,ZhangNCA2023} and game character control~\cite{MaAIIDE17,Jansen2008DirectionMF}.


Many algorithms have been proposed to solve LMAPF. The \textit{replan-based} algorithms~\cite{MaAAMAS17,LiAAMAS20a,KouAAAI20,DamaniRAL21} decompose LMAPF into a series of MAPF problems and solve them sequentially. 
Although replan-based algorithms possess high solution quality for small problems, they scale poorly to real-world scenarios with large maps, large numbers of agents, and limited planning time. 

\begin{figure}[!t]
    \centering
    \includegraphics[width=0.45\textwidth]{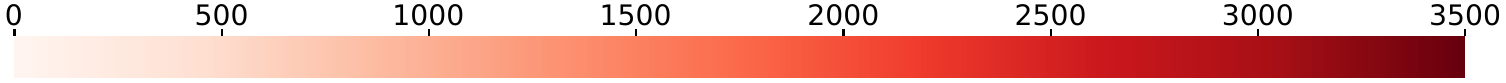}
    \subfigure[No guidance.]{
        \includegraphics[width=0.14\textwidth]{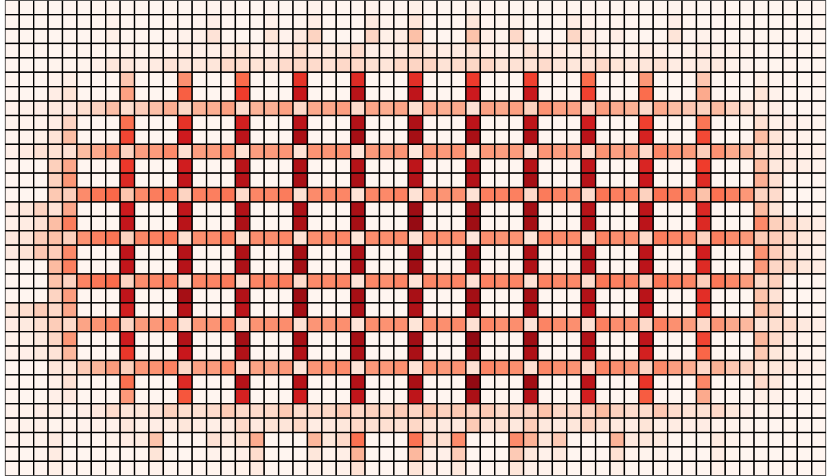}
    }
    \subfigure[Offline guidance.]{
      \includegraphics[width=0.14\textwidth]{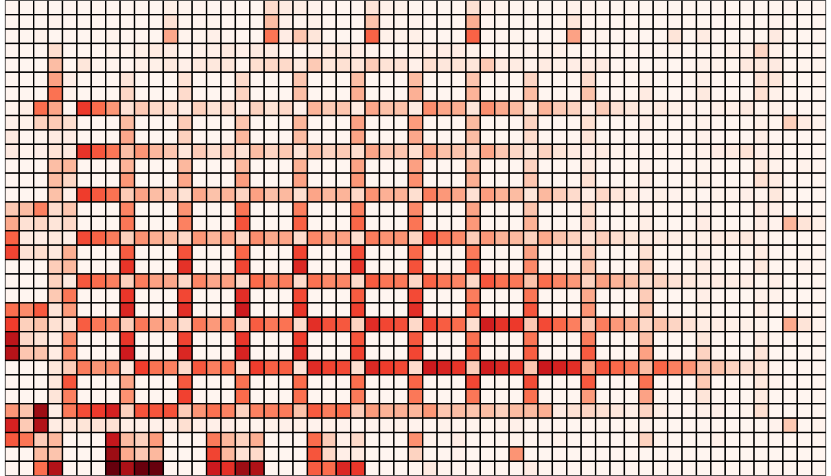}
    }
    \subfigure[Online guidance.]{
      \includegraphics[width=0.14\textwidth]{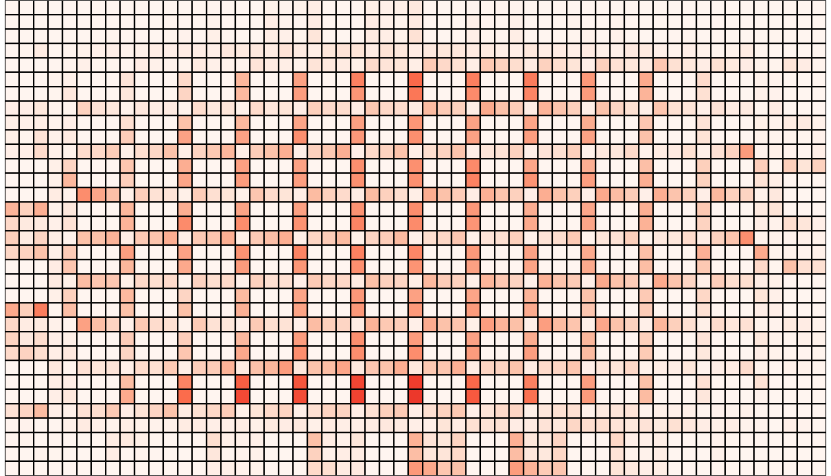}
    }
    \caption{
    Comparison of no guidance, offline guidance \cite{zhang2024ggo}, and our online guidance with a simulation of 5,000 timesteps with 600 agents in a warehouse map of size 33 $\times$ 57 with 1,091 non-obstacle cells. The average and standard deviation of throughput over 10 simulations for no guidance, offline guidance, and online guidance are $3.18 \pm 0.04$, $6.42 \pm 0.09$, and $8.66 \pm 0.04$, respectively. The heatmaps show the number of times the agents take wait action in each cell, approximating the level of congestion. Our online guidance results in the most balanced traffic and thus less congestion and higher throughput.
    }
    \label{fig:front-fig}
\end{figure}

Meanwhile, the \textit{rule-based} algorithms~\cite{WangICAPS08,okumura2019priority,Yu2023} compute a shortest path or a shortest-distance heuristic for each agent without considering the collisions and use pre-defined rules to resolve the collisions. Rule-based algorithms run very fast and scale to extremely large maps and large numbers of agents.

However, rule-based algorithms have no guarantee on the solution quality. \citet{zhang2024ggo} have shown that with 150 agents, the throughput of RHCR, a state-of-the-art \textit{replan-based} algorithm, is 24.2\% better than PIBT, a state-of-the-art \textit{rule-based} algorithm, in a 33 $\times$ 36 warehouse map. To improve the solution quality of rule-based algorithms, prior works~\cite{zhang2024ggo,ChenAAAI24,Yu2023,li2023study} have explored providing \textit{guidance} to the agents such that they automatically avoid congested areas, thereby improving throughput.
Most previous works adopt a static \textit{offline} guidance, with the \textit{guidance graph}~\cite{zhang2024ggo} providing a versatile and state-of-the-art representation of guidance. 
Upon optimizing an offline guidance for PIBT, the throughput gap between PIBT and RHCR is reduced to less than 4.2\%~\cite{zhang2024ggo}.
Nevertheless, the offline nature of the guidance graph assumes that the area of congestion does not change, an assumption 
that does not always hold in the real world.
For example, in automated warehouses, task distribution can shift because of changes in the distribution of orders.

Therefore, instead of optimizing an offline guidance graph that provides static guidance, we optimize an \emph{online guidance policy} capable of adapting a dynamic guidance graph over time based on real-time traffic information to improve the throughput of PIBT-based LMAPF algorithms, the state-of-the-art in rule-based LMAPF. 
\Cref{fig:front-fig} shows the traffic congestion resulting from different guidance.

To incorporate the online guidance policy into rule-based LMAPF algorithms, we design two pipelines. One directly uses the guidance policy to dynamically generate guidance graphs based on real-time traffic information and replaces uniform edge weights with the generated guidance graphs. The other uses the dynamically generated guidance graphs to adaptively plan better guide paths and move the agents along the guide paths while resolving collisions. To optimize the guidance policy, we follow \citet{zhang2024ggo} in using Covariance Matrix Adaptation Evolutionary Strategy (CMA-ES)~\cite{hansen2016cmaes}, a single-objective black-box optimization algorithm, to optimize the policy.

We make the following contributions:
(1) We generalize the offline, static guidance graph to an online, dynamic guidance policy capable of updating the guidance graph based on real-time traffic information.
(2) We propose two methods to incorporate guidance policy into PIBT.
(3) We address the issue of dynamic task distribution, a critical concern in industrial settings. Our online guidance policy was tested in LMAPF simulations with both static and dynamic task distributions. The results demonstrate its advantages over offline guidance with an improvement in throughput of up to 30.75\%, and over human-designed online guidance with an improvement of up to 52.42\% across various maps and algorithms.


\section{Preliminary and Related Work}
\label{sec:preliminary}
\subsection{LMAPF and LMAPF Algorithms}
\begin{definition}[Lifelong MAPF (LMAPF)]
Given a graph $G(V, E)$, a set of agents and their corresponding start and goal locations, LMAPF aims to move all agents from their start to their goal locations while avoiding collisions.
Furthermore, agents are constantly assigned new goals upon reaching their current goals. 
At each timestep, agents may either move to an adjacent vertex or remain stationary. Collisions happen when two agents are at the same vertex or swap locations in the same timestep. 
The objective is to maximize throughput, defined as the average number of goals reached per timestep.
\end{definition}
LMAPF algorithms mainly include \emph{replan-based} algorithms~\cite{Li2020LifelongMP,LiuAAMAS19,Li2020LifelongMP} and \emph{rule-based} algorithms~\cite{WangICAPS08,okumura2019priority,Yu2023}. Replan-based algorithms decompose the LMAPF problem into a sequence of MAPF problems and rely on MAPF algorithms to solve them.  RHCR~\cite{Li2020LifelongMP}, Rolling-Horizon Collision Resolution, is the state-of-the-art of this category. For every $h$ timesteps, RHCR replans collision-free paths of all agents within a time window of $w$ timesteps ($w \geq h$), ignoring collisions beyond it. 
While RHCR has state-of-the-art throughput with a small number of agents in small graphs (or ``maps'' as we will refer to them later) ~\cite{Li2020LifelongMP}, it scales poorly to instances with large numbers of agents and limited planning time. Prior works have shown that the throughput of RHCR drops to almost zero with more than 200 agents in a 33 $\times$ 36 small warehouse~\cite{ZhangNCA2023,zhangLayout23} with a per-5-timestep planning time limit of 60 seconds. Even with optimized guidance graph, \citet{zhang2024ggo} shows that RHCR does not scale to more than 250 agents in the same map. 

Rule-based algorithms, on the other hand, run much faster than replan-based ones at the expense of solution quality. Rule-based algorithms compute a shortest path or a shortest distance heuristic for each agent without considering the collisions and leverage pre-defined rules to move the agents while resolving collisions. PIBT~\cite{okumura2019priority}, Priority Inheritance with Backtracking, is the state-of-the-art of this category. \citet{ChenAAAI24} shows that the planning time of PIBT per timestep is over 150 times faster than RHCR. \citet{Jiang2024Competition} further shows that PIBT is currently the only existing algorithm that can handle a limited planning time of $1$ second for up to $10,000$ agents.
Therefore, we focus on improving the throughput of PIBT using guidance.

\mysubsubsection{PIBT. } 
PIBT maintains and adjusts a priority for each agent and leverages a single-timestep rule to move agents. At each timestep, PIBT ranks each agent's actions based on the shortest distance from its next location to its goal, with a preference for shorter distances. By default, an agent always tries to follow its shortest path if no higher-priority agents act as obstacles. Otherwise, a lower-priority agent needs to avoid collisions with higher-priority agents by taking the next preferred action. If a lower-priority agent fails to avoid a collision, PIBT applies a backtracking mechanism that forces higher-priority agents to take their next preferred actions and retries all the movement until all agents can take a collision-free action. Note that PIBT possesses favorable properties; for instance, it is complete for LMAPF on bi-connected graphs.

\subsection{Guidance in MAPF and LMAPF}

While not necessarily using the term ``guidance'', the general idea of guiding agents' movements is widely explored. Prior works fall into two categories, namely \textit{offline} guidance that provides static guidance to the agents, and \textit{online} guidance in which the guidance changes over time based on real-time traffic information.

\mysubsubsection{Offline Guidance.}
A static offline guidance usually leverages edge directions or edge costs to encourage the agents to move along certain edges. The pre-defined directions and costs are generated in advance and are not updated during the execution of the algorithm.
\citet{WangICAPS08} leverages unidirectional edges in a graph to force the agents to move in one direction, eliminating head-to-head collisions. The idea of highway~\cite{Cohen2015FeasibilitySU,liron_highway16,lironPhDthesis,li2023study} pre-defines a subset of edges in the graph as highway-edges and encourages the agents to move along those edges. 

The state-of-the-art offline guidance is the guidance graph~\cite{zhang2024ggo}. It unifies the representation of guidance by using a directed weighted graph. The edge weights of the guidance graph define the cost of moving and waiting for agents at different locations. The edge weights of the guidance graphs are optimized automatically either directly using a single-objective derivative-free optimizer CMA-ES~\cite{hansen2016cmaes} or indirectly with Parameterized Iterative Update (PIU), an iterative update procedure. Starting with a uniform guidance graph with all edge weights being 1, PIU runs the LMAPF simulator to collect the average traffic information, which is used by a parameterized update model to update the guidance graph. The updated model is then optimized by CMA-ES for PIU to generate high-throughput guidance graphs.

In this work,  we adopt the definition of guidance graphs from previous work~\cite{zhang2024ggo}, with slight rephrasing.

\begin{definition}[Guidance Graph]
\label{def:gg}
Given a graph $G(V, E)$, a guidance graph is a directed weighted graph $G_g(V_g, E_g,\omegaV)$, where $V_g=V$, and $E_g = \{(u, v) | u, v\in V\}$ encodes the action costs in every single vertex, including moving and waiting, for all agents. The action costs are represented collectively as the edge weights $\omegaV \in \mathbb{R}^{|E_g|}_{>0}$.
\end{definition}

\mysubsubsection{Online Guidance.}
Online guidance is able to adapt the guidance based on real-time traffic information during the execution of the LMAPF algorithm.
\citet{Jansen2008DirectionMF} assigns a direction vector to each location in the graph and encourages the agents to move along the direction vector by setting the movement cost of that location to be inversely related to the dot product of the direction vector and the edge cost. The direction vectors are computed from the past traffic information by using a handcrafted equation. \citet{han2022spaceutil} computes a handcrafted temporal heuristic function to estimate the movement cost, guiding the agents' movement. Similarly, \citet{ChenAAAI24} and \citet{learntofollow2024} collect the planned paths of all agents and use handcrafted equations to compute the movement costs.
The above methods rely solely on handcrafted functions, which demand significant human effort and greatly limit the effectiveness of online guidance.
\citet{Yu2023} relies on a trained data-driven model to predict the movement delays of the agents and uses the delays as the movement costs. However, predicting delays is not directly related to throughput, which is the objective of LMAPF.

\subsection{Challenge of Online Guidance}
Since online guidance is able to adapt the guidance on-the-fly based on real-time traffic information, it should be able to provide better guidance for agents when the traffic pattern shifts abruptly. For example, in automated warehouses in which robots are used to transport packages from one location to another, the distribution of packages might change over time, resulting in different traffic patterns. In this case, it is necessary to provide online guidance to agents.

However, it is non-trivial to incorporate online guidance into LMAPF algorithms because of the computational overhead of computing the heuristic values. And it is unavoidable because PIBT requires the cost-to-go heuristic, which is the shortest distance from the current locations to the goal locations, to plan agents' actions. With an offline guidance graph, the heuristic values only need to be computed once at the beginning. With an online guidance graph, however, the heuristic values of the entire map need to be updated once the guidance graph is updated. To tackle this issue, a recent work~\cite{ChenAAAI24} proposes a variant of PIBT, referred to as Guided-PIBT (GPIBT).

\mysubsubsection{GPIBT.} GPIBT first plans a guide path for each agent that minimizes a handcrafted congestion cost equation, which takes other agents' guide-path edge usage into account.
It then moves agents to their goals following the guide paths and resolves collisions using PIBT. Since the heuristics that guide agents to follow guide paths only need to be updated while agents are assigned new goals, GPIBT reduces the computational overhead of heuristic computation. However, since the guide paths are not updated until the agents reach their current goals, GPIBT is less flexible than PIBT in terms of the paths that the agents eventually take to reach the goals.

Given the state-of-the-art performance of PIBT and GPIBT, we focus on incorporating automatically optimized online guidance in them, thereby improving their performance in our work.

\section{Approach} \label{sec:approach}
In this section, we first introduce the policy for generating dynamic guidance, defined in \Cref{def:on-gg}. We then discuss how guidance policies are integrated with PIBT-based LMAPF algorithms in \Cref{sec:application}. Finally, we provide a detailed explanation of the optimization of the guidance policy in \Cref{sec:optimization}.

\begin{definition}[Guidance Policy]
\label{def:on-gg}
Given a guidance graph $G_g(V_g, E_g, \omegaV)$, a guidance policy is a function $\pi_{\thetaV}: \OV \rightarrow \mathbb{R}^{|E_g|}_{>0}$ that computes the updated edge weights $\omegaV' \in \mathbb{R}^{|E_g|}_{>0}$ given the observation $\oV \in \OV$ collected in a LMAPF simulation. The policy $\pi_{\thetaV}$ is parameterized by $\thetaV \in \ThetaV$, where $\ThetaV$ is the space of all parameters.
\end{definition}

Depending on the LMAPF algorithm, the observation of guidance policy consists of one or more past traffic patterns, the current distribution of tasks, and future planned paths. We discuss the choice of observation in \Cref{sec:application}.

\subsection{Incorporating Guidance Policy}
\label{sec:application}

Here, we discuss two ways of using guidance graphs and guidance policies in LMAPF algorithms.

\begin{figure}[!t]
    \centering
    \includegraphics[width=0.8\linewidth]{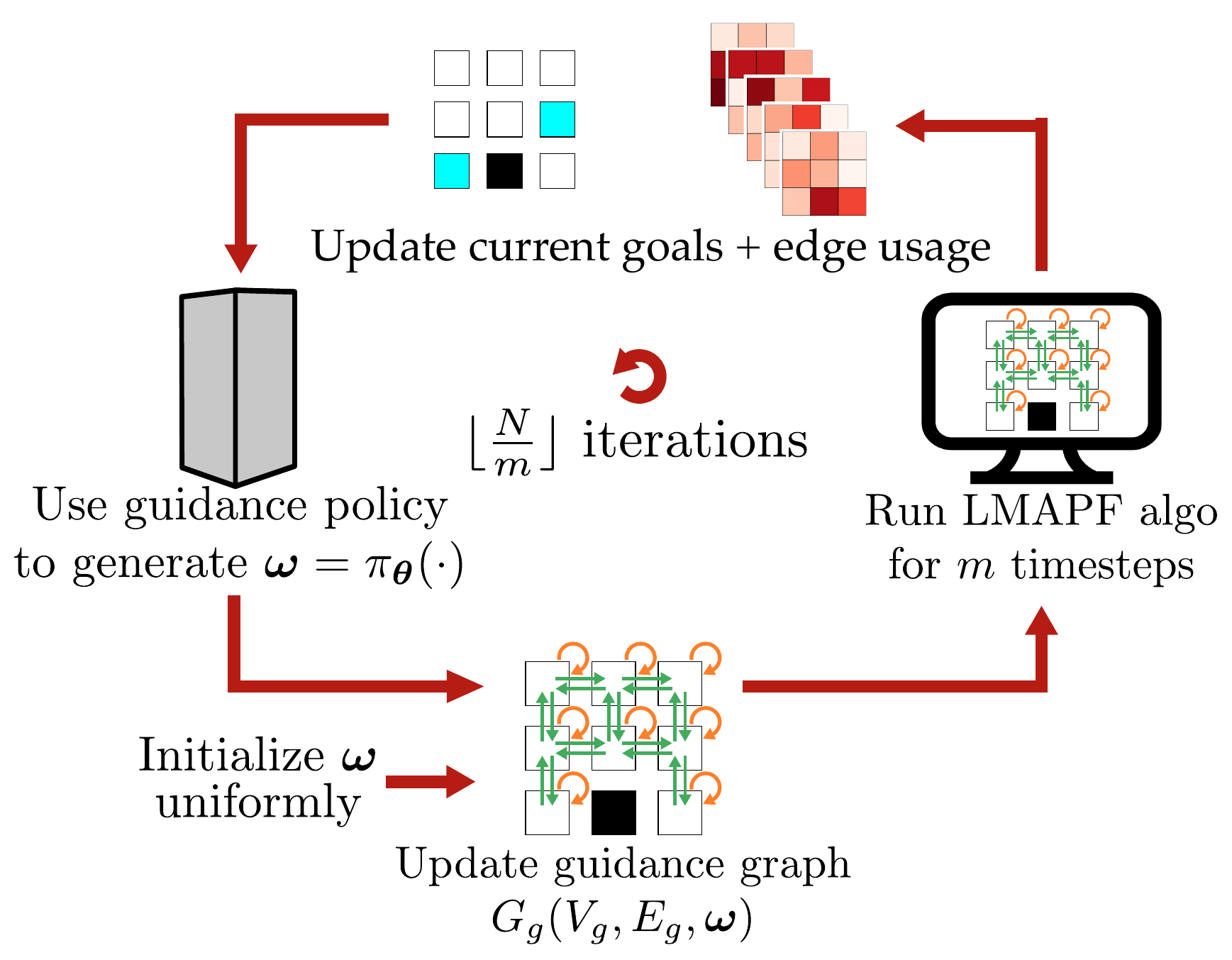}
    \caption{Overview of incorporating guidance policy with Direct Planning algorithms like PIBT.
    }
    \label{fig:pipeline-pibt}
\end{figure}
\mysubsubsection{Direct Planning. }
Most LMAPF algorithms directly plan on the guidance graph by minimizing the sum of action costs encoded in the guidance graph. PIBT~\cite{okumura2019priority} falls into this category. 
\Cref{fig:pipeline-pibt} shows the overview of this method. Starting from a uniform guidance graph with all weights being 1, we run the LMAPF algorithm for $m$ timesteps. After $m$ timesteps, we obtain the \emph{edge usage} (i.e., the number of timesteps that each edge in $E_g$ is used by some agent) of the past $m$ timesteps and the currently assigned goals of the agents. The edge usage reflects the recent traffic information, and the current goals project the future possible congestion locations. Upon obtaining the edge usage and current goals, we use the guidance policy to update the guidance graph, starting a new iteration. 

Since we end the LMAPF simulation after $N$ timesteps, we run the above procedure for $\lfloor \frac{N}{m} \rfloor$ iterations. By choosing $m$, we control the trade-off between the adaptability of the online updated guidance graph and additional computational costs associated with the online update mechanism.
This is because we need to maintain a heuristic table that contains the shortest path length between every pair of vertices on the guidance graph, which is used for PIBT to rank the actions for every agent at every timestep. 
We make a slight change here: instead of PIBT ranking neighbors by the shortest distance from each neighbor to the goal location, we follow \citet{zhang2024ggo} and rank the neighbors by the sum of the action cost from the current location to the neighbors (including itself) and the distance from the neighbors to the goal. That is because, with the guidance graph, the cost from the current location to the neighbors varies. Additionally, this approach allows agents to choose to stay at their current location due to the existence of self-edges on the guidance graph.
Admittedly, the heuristics table needs to be updated every time the guidance graph is updated, thereby making the LMAPF algorithm more computationally expensive. 
During implementation, we use techniques to reduce recomputation costs. Since the guidance graph updates frequently, calculating the shortest paths for all vertex pairs often results in unnecessary computations. Within one update interval, agents have a limited number of goal locations, and it is likely that some locations on the map are never visited by any agents. Instead of computing the shortest distances for all vertex pairs, we refer \cite{SilverAIIDE05} and use a lazy mechanism. Each agent maintains a search tree rooted at its goal location. When the shortest distance at a location is queried, the tree expands until it finds this location, saving a lot of unnecessary computation.

\begin{figure}[!h]
    \centering
    \includegraphics[width=0.8\linewidth]{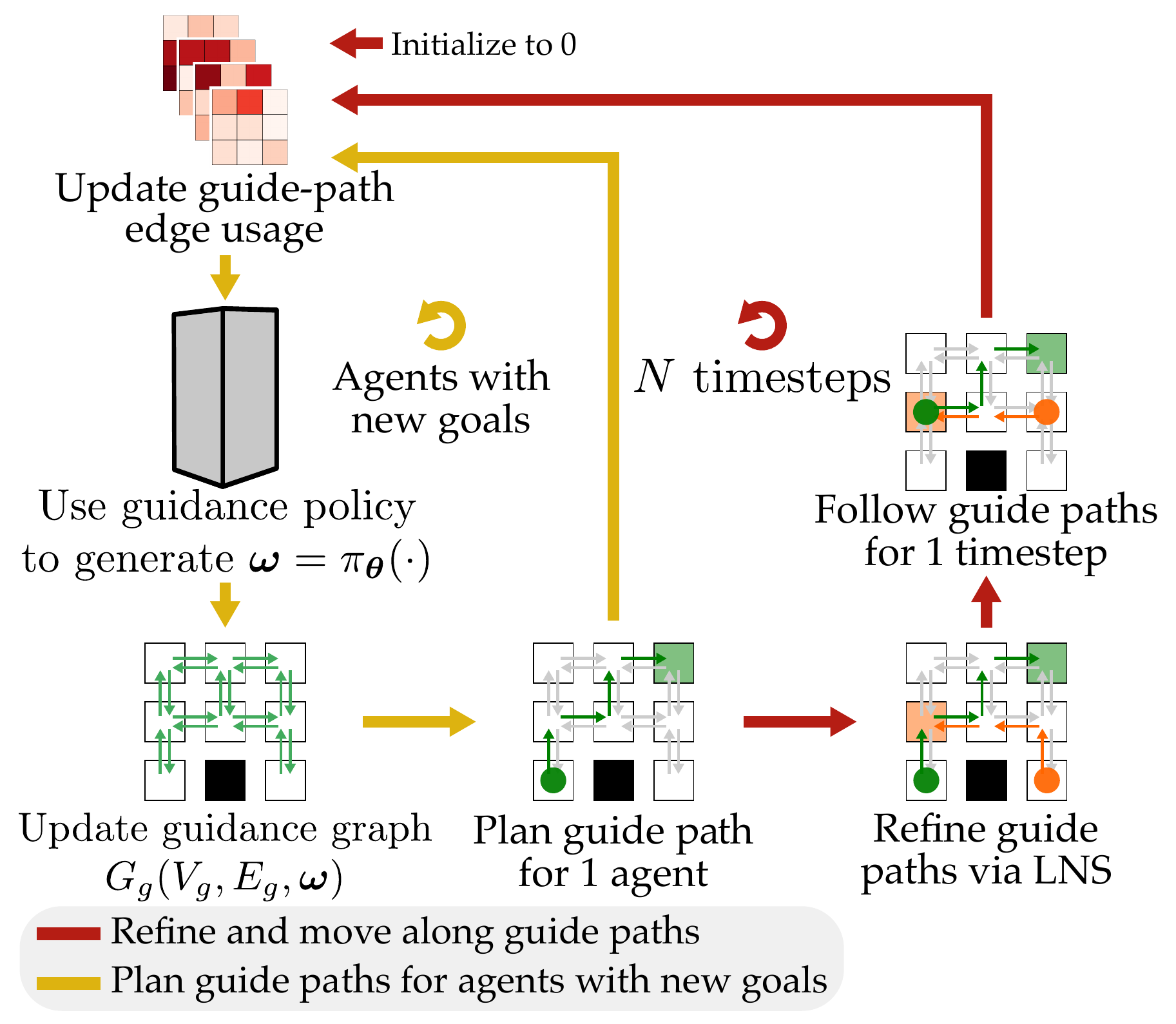}
    \caption{Overview of incorporating guidance policy with Guide-Path Planning algorithms like GPIBT. 
    }
    \label{fig:pipeline-gpibt}
\end{figure}
\mysubsubsection{Guide-Path Planning. } 

Given the heavy computational overhead of repeatedly computing heuristic tables or trees, another approach is to use guide paths. \Cref{fig:pipeline-gpibt} shows the overview. 

We generally follow the GPIBT framework~\cite{ChenAAAI24} but modify the guide-path generation process. At each timestep, GPIBT plans guide paths for each agent assigned a new goal in a sequential manner (yellow loop). It may then use Large Neighborhood Search (LNS)~\cite{ijcai2021LNS} to refine the guide paths for some randomly selected agents. Finally, each agent moves in a PIBT style, ranking its neighbors according to the distance on the original graph (instead of the guidance graph) to their guide paths (red loop). This procedure is repeated for $N$ timesteps.

We incorporate our guidance policy during guide-path generation. The guide paths for agents are the shortest paths from their previous goal locations to their current goal locations on the guidance graph. After planning guide paths for each agent, the guidance graph updates. GPIBT originally uses a handcrafted equation to compute the guidance graph weights based on all agents' guide-path edge usage. In our method, we replace this equation with our guidance policy.

The guide-path edge usage is initialized to 0 and updated whenever any agent's guide path changes. This approach combines past and future traffic information, as each agent follows its guide path from its previous goal to its current location and will continue following its guide path until the agent reaches its current goal (with possible temporary deviations to avoid collisions), at which point a new guide path is generated to guide it to its next goal.

Note that for GPIBT, the guidance graph is only used to compute the guide paths, and there is never a ``wait'' action on the guide paths. Therefore, there is no need to incorporate self-edges on the guidance graph.

Compared to direct planning, guide-path planning minimizes the effort on heuristic updates because it does not maintain the pairwise distances between vertices. It only computes the agents' guide paths once when an agent is assigned a goal. If agents deviate from their guide paths, they do not plan new paths but try to return to the guide paths, which is much less computationally intensive.

\subsection{Guidance Policy Optimization}
\label{sec:optimization}
We aim to optimize the parameter $\thetaV$ of the guidance policy $\pi_\theta$ to maximize the throughput given by the LMAPF simulators. Following \citet{zhang2024ggo}, we use CMA-ES~\cite{hansen2016cmaes}, a single-objective derivative-free optimization algorithm, to optimize $\thetaV$. 
CMA-ES maintains a multi-variate Gaussian distribution to represent the distribution of solutions. Starting with a standard normal Gaussian, CMA-ES samples a batch of $b$ parameter vectors $\thetaV$, forming $b$ guidance policies. It then evaluates these guidance policies by running the given LMAPF simulator $N_e$ times to compute the average throughput. Finally, based on the evaluated guidance policies, CMA-ES updates the parameters of the Gaussian towards the high-throughput region of the search space. CMA-ES repeats the above procedure until the number of guidance policy evaluations reaches $N_{eval}$. 
We include information on the selection of CMA-ES-related hyperparameters in \Cref{appendix:exp:hyper}.

Note that if the dimension of $\thetaV$ is too large, it becomes difficult for CMA-ES to optimize. Therefore, we use a Convolutional Neural Network(CNN) for PIBT with 3,119 parameters and a specialized windowed quadratic network with 560 parameters. Detailed information can be found in \Cref{appendix:exp:on-pibt-gp,appendix:exp:on-gpibt-gp}. Admittedly, the design of our current guidance policy architecture can only handle 4-neighbor grids instead of general graphs. However, 4-neighbor grids are currently the main focus of the community. Moreover, the concept of applying the guidance policy is not limited to the current network structures.


\section{Experiments and Analysis} \label{sec:experiment}
\subsection{Experiment Setup}
\label{sec:exp-overview}

\begin{figure*}[!t]
    \centering
    \includegraphics[width=0.9\linewidth]{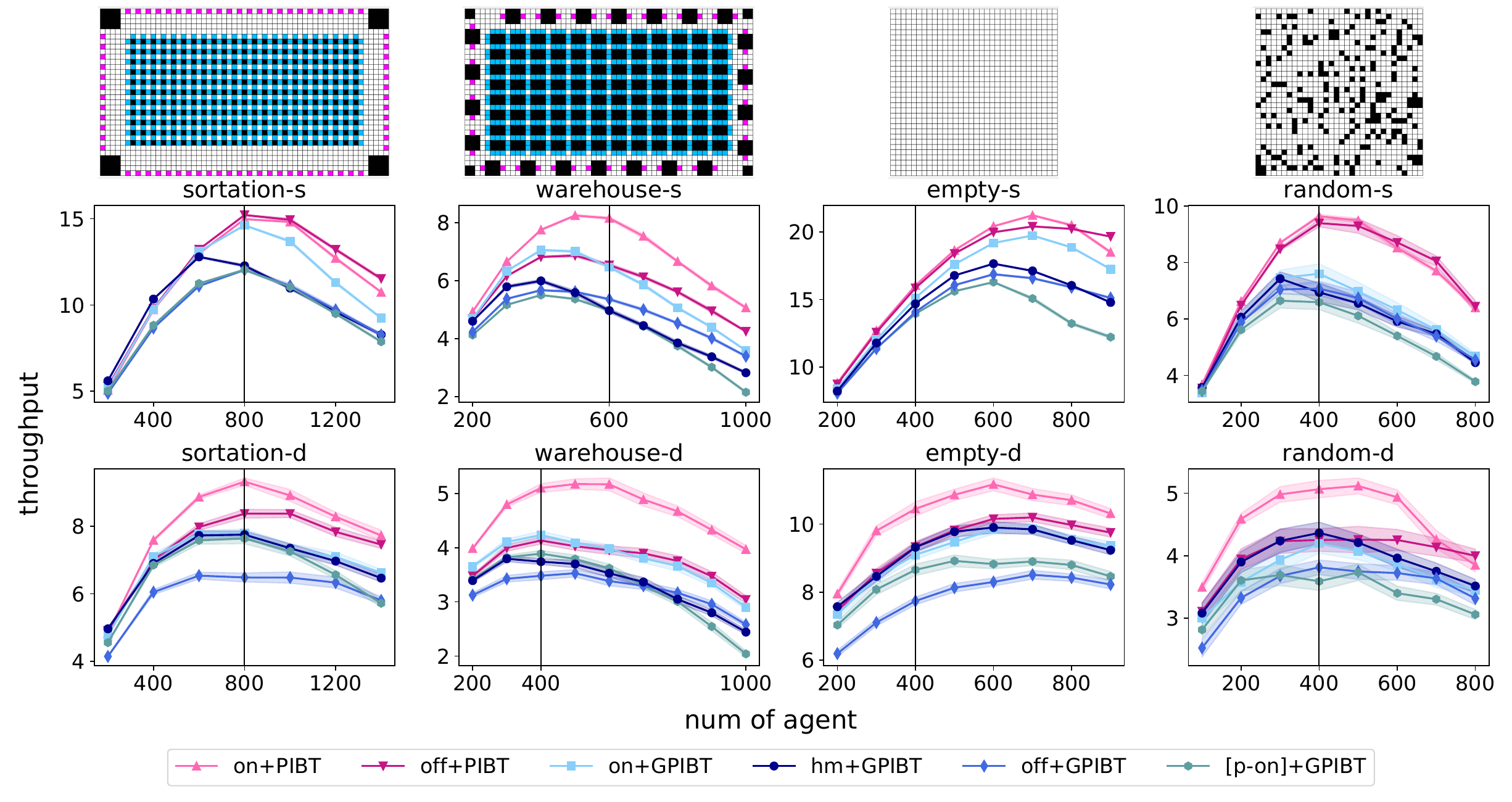}
    \caption{Throughput with different numbers of agents. The black vertical lines show the number of agents that are used to optimize the guidance policies. The solid line shows the average throughput over 50 LMAPF simulations, and the shaded areas denote the 95\% confidence interval. ``s'' and ``d'' stand for static and dynamic task distribution, respectively. }
    \label{fig:all}
\end{figure*}
We conduct experiments comparing online and offline guidance, optimized guidance policies versus human-designed guidance policies, and the advantages of online guidance with dynamic task distribution. We also compare the runtime of all algorithms. Then, we present results for guidance policies for GPIBT with LNS. Furthermore, we demonstrate that our approach mitigates deadlock issues in PIBT-based algorithms.
 
\mysubsubsection{Maps.}
We conduct experiments on 4 maps: (1) \textit{sortation-33-57}, (2) \textit{warehouse-33-57}, (3) \textit{empty-32-32}, and (4) \textit{random-32-32}, shown on top of \Cref{fig:all}.
The first two have regular patterns and are used to test MAPF algorithms in automated warehouse settings. Specifically, the \textit{sortation} map is the same as in \cite{ChenAAAI24} and the \textit{warehouse} map is generated by us. 
The latter two are selected from the MAPF benchmark~\cite{SternSoCS19}.

\mysubsubsection{Task Generation.}
As shown at the top of \Cref{fig:all}, the \textit{sortation} and \textit{warehouse} maps have workstations (pink) and endpoints (blue). Each agent's goals constantly alternate between the workstations and the endpoints, simulating the warehousing scenario in which robots pick up items from the workstations and drop them off in chutes or shelves (black) reachable from the endpoints in the middle.
For \textit{empty} and \textit{random} maps, goals are sampled from empty spaces (white). 

\mysubsubsection{Task Distribution.}
We consider both static and dynamic task distributions. Static task distribution samples goals uniformly, the most common setting in previous works~\cite{ChenAAAI24,zhangLayout23,ZhangNCA2023}. For the dynamic task distribution, goals are sampled from the Gaussian or multi-modal Gaussian distribution, where the Gaussian centers change for every 200 timesteps. Concretely, in \textit{sortation} and \textit{warehouse} maps, goals on endpoints are sampled from a Gaussian distribution, and goals on workstations are sampled uniformly. For \textit{empty} and \textit{random} maps, goals are sampled from a multi-modal Gaussian distribution with $K$ Gaussian centers. The hyperparameters for the distribution are provided in \Cref{tab:dist-hyper} in \Cref{appendix:exp:hyper}.

\mysubsubsection{Algorithm Comparison.}
We compare 6 algorithms, each with a different LMAPF solver or guidance approach. 
\begin{description}
    \item[on+PIBT] our proposed online guidance policy applied to PIBT. We optimize an online guidance policy, periodically call it to update the guidance graph, and directly plan on the guidance graph.
    \item[off+PIBT] PIBT with an offline guidance graph, optimized using PIU~\cite{zhang2024ggo}. 
    \item[on+GPIBT] our proposed online guidance policy applied to GPIBT. We optimize an online guidance policy and call it when each agent plans its guide path.
    \item[\text{[p-on]}+GPIBT] GPIBT with periodically updated guidance graph. We use the same pipeline as on+PIBT. The inputs of online policy are past traffic and current goals instead of guide-path information. We present the results of this algorithm to show the advantages of on+GPIBT. 
    \item[off+GPIBT] GPIBT with an offline guidance graph, optimized using PIU~\cite{zhang2024ggo}. The guide paths are generated according to the optimized offline guidance graph.
    \item[hm+GPIBT] the original GPIBT that uses a human-designed equation as the guidance policy~\cite{ChenAAAI24}. Specifically, we compute $\omegaV$ using the SUM\_OVC function~\cite{ChenAAAI24}, where each weight is the sum of the guide-path vertex usage and the product of the head-on-head guide-path edge usage. 
\end{description}

Note that for the PIBT-based algorithms, we used the ``swap'' technique mentioned in \cite{okumura2023lacam2}. This is necessary because vanilla PIBT is only deadlock-free on bi-directed graphs, and the \textit{random} map does not meet this criterion. With the ``swap'' technique, the deadlock issue can be mitigated a lot. Additionally, unless specifically mentioned, we do not include LNS in the main results for all GPIBT-based algorithms, which limits their performance. In one of the following subsections, we present the results with LNS included.

We run all LMAPF algorithms for $N = 1,000$ timesteps. In on+PIBT and [p-on]+GPIBT, we update the guidance graph for every $m=20$ timesteps. The CPU runtime for all algorithms is measured on a local machine with a 64-core AMD Ryzen Threadripper 3990X CPU, 192 GB of RAM, and an Nvidia RTX 3090Ti GPU. More compute resource information can be found in \Cref{appendix:exp:compute}. For relevant software libraries, see \Cref{appendix:exp:software}.

\subsection{Experiment Results} \label{subsec:exp-result}

\Cref{fig:all} compares the throughput of different algorithms. To show the generalizability of our approach, we optimize the guidance policies and guidance graphs with the numbers of agents indicated by the black vertical lines. We then evaluate them with various numbers of agents. We highlight several key findings below.

\mysubsubsection{Online vs. Offline Guidance. }
\label{sec:on-and-off}
We first compare our online guidance policy with the offline guidance graph~\cite{zhang2024ggo}. That is, we focus on comparisons between on+PIBT with off+PIBT and on+GPIBT with off+GPIBT in \Cref{fig:all}. Under most settings, on+PIBT and on+GPIBT generally match or outperform off+PIBT and off+GPIBT throughout different numbers of agents, respectively.  However, in the \textit{sortation} map with static task distribution, on+PIBT is (slightly) worse than off+PIBT because congested locations do not change a lot over time. Therefore, a well-optimized offline guidance graph can alleviate congestion well enough. 
Besides, there exist outliers on some number of agents because our model is not directly optimized for these cases.
    

\mysubsubsection{Static vs. Dynamic Task Distribution. }
\label{sec:dyn-dist}
We expect that dynamic task distribution gives more advantages to online guidance over offline guidance because the guidance policy can dynamically change the guidance graph based on real-time traffic information, which depends on the task distribution. 
Clearly in \Cref{fig:all}, by comparing on+PIBT with off+PIBT, our online guidance policy is more advantageous than the offline guidance graph when moving from static to dynamic task distributions. However, the improvement ratio of on+GPIBT over off+GPIBT does not consistently increase when moving from static to dynamic task distributions. This is because the difference between on+GPIBT and off+GPIBT goes beyond simply online vs. offline: on+GPIBT generates the guidance graph based on the current guide paths, meaning that a new guide path is more likely to avoid areas frequently used by the guide paths of other agents, while off-GPIBT uses a static guidance graph, providing no incentive to generate guide paths that avoid congestion with other guide paths. To isolate the impact of the online mechanism alone, we compare [p-on]+GPIBT against off+GPIBT, where neither algorithm uses guide-path information to generate guidance graphs. In this case, the improvement ratio consistently increases while moving from static to dynamic task distributions. In addition, the comparison between on+GPIBT with [p-on]+GPIBT shows the effectiveness of using guide-path information. 

\mysubsubsection{Optimized vs. Handcrafted Online Guidance. }
\label{sec:learn-or-not}
We also compare our optimized online guidance policy (on+GPIBT) with the human-designed guidance policy (hm+GPIBT). As shown in \Cref{fig:all}, on+GPIBT significantly outperforms hm+GPIBT under a static task distribution. For a dynamic task distribution, both on+GPIBT and hm+GPIBT capture the traffic patterns effectively, resulting in similar performance. In such dynamic settings, we believe that hm+GPIBT is near optimal, considering the observations, task distribution, and the map. To further illustrate this, we optimize the on+GPIBT policy with fewer parameters. The detailed explanation and results are provided in \Cref{appendix:hm}. However, for the \textit{warehouse} map, on+GPIBT maintains an advantage over hm+GPIBT.

\mysubsubsection{Runtime. }
Admittedly, online mechanisms incur more computational overhead.
on+PIBT is up to 4 times slower than off+PIBT due to the need to recompute the heuristic tables. on+GPIBT is up to 7 times slower than hm+GPIBT and off+GPIBT due to the need to update the guidance graph every time an agent replans its guide path.
However, this runtime overhead is often acceptable in practice, as the average runtime per timestep never exceeds 0.026 seconds across all experiments for all algorithms, which is fast enough for realistic settings. 
Besides, on average, on+PIBT algorithms are two to three times slower than on+GPIBT algorithms because on+GPIBT uses the guide-path mechanism to reduce the computational overhead of heuristics. Numerical results for runtime can be found in \Cref{tab:runtime} in \Cref{appendix:runtime}.

\begin{figure}[!h]
    \centering
    \includegraphics[width=\linewidth]{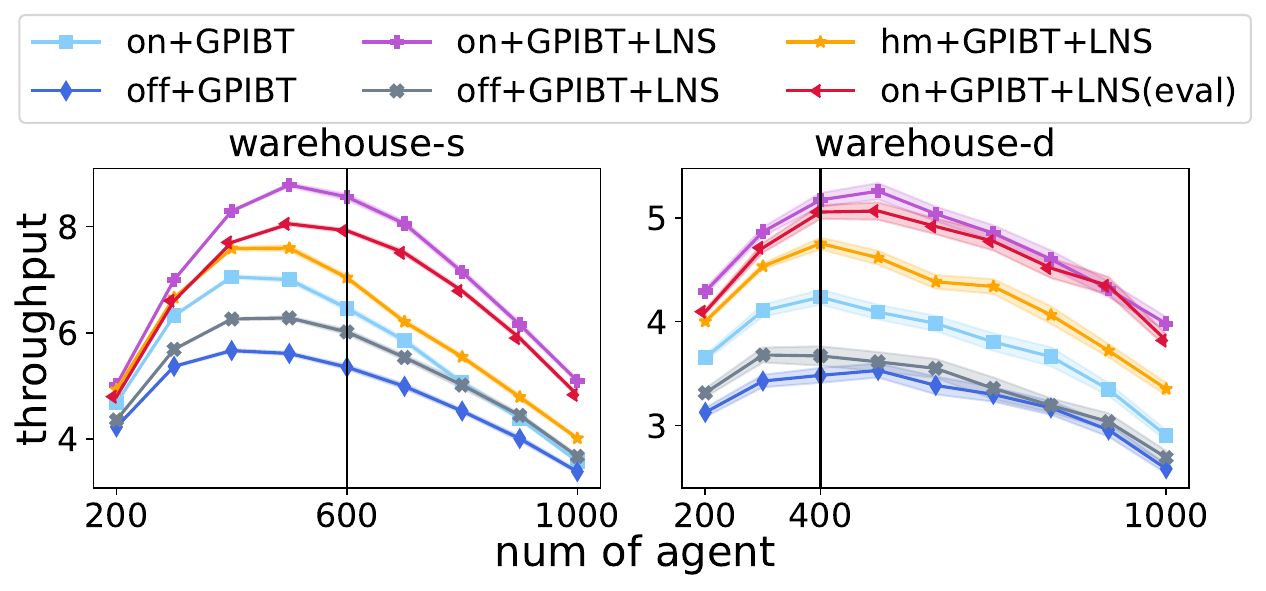}
    \caption{GPIBT with LNS results. The notation of this figure is similar to that in \Cref{fig:all}. }
    \label{fig:lns}
\end{figure}
\mysubsubsection{GPIBT with LNS. } \Cref{sec:approach} mentioned that the GPIBT-based methods can incorporate LNS to improve the quality of guide paths and further enhance throughput. LNS refines agents' guide paths with an iterative procedure. In each LNS iteration, $n_g$ agents are selected to replan their guide paths from their current locations. If the new sum of all agents' guide-path lengths on the guidance graph is smaller, LNS accepts the new guide paths. At each timestep, LNS is terminated either if it exceeds $N_{iter}$ iterations or if it runs for more than $t_{LNS}$ seconds.
We focus on the \textit{warehouse} map with both static and dynamic task distribution. LNS parameters are set as $N_{iter}=10$, $n_g=10$, and $t_{LNS}=8$. CMA-ES hyperparameters are kept the same as in the main results.

\Cref{fig:lns} shows the \textit{throughput-agents} curve with LNS experiments. The results include on+GPIBT+LNS, off+GPIBT+LNS, hm+GPIBT+LNS, as well as on+GPIBT+LNS(eval). We optimize the first three guidance policies or guidance graphs with LNS included in the optimization loop. For on+GPIBT+LNS(eval), we use the on+GPIBT guidance policy and incorporate LNS only during the evaluation phase.

Comparing on+GPIBT with on+GPIBT+LNS and on+GPIBT+LNS(eval) shows that LNS substantially improves the throughput. The throughputs of on+GPIBT+LNS, on+GPIBT+LNS(eval), and the hm+GPIBT+LNS all dominate that of off+GPIBT+LNS, indicating that under the LNS setting, all online methods are better than the offline method. The on+GPIBT+LNS(eval) policy ranks as the second best, only outperformed by the on+GPIBT+LNS approach. This result demonstrates the generalizability of our guidance policy, which is capable of sidestepping the costly optimization of guidance policies with LNS. Interestingly, the improvement of on+GPIBT over off+GPIBT is larger than off+GPIBT+LNS over off+GPIBT, indicating that the online mechanism is more helpful than LNS. That is because off+GPIBT ignores other agents' guide-paths. LNS only helps when an agent deviates from its guide path, replanning the agent's guide path from the deviated location instead of continuing to follow the previously planned guide path. However, ignoring other agents' guide paths cannot lead to higher throughput. 

As shown in \Cref{tab:runtime} and \Cref{tab:lns-runtime} in \Cref{appendix:runtime}, all LNS-based algorithms are significantly slower than those without LNS. However, the runtime remains acceptable, with the most expensive algorithm taking only 0.043 seconds per timestep.

\begin{figure}[h]
    \centering
    \includegraphics[height=0.8in]{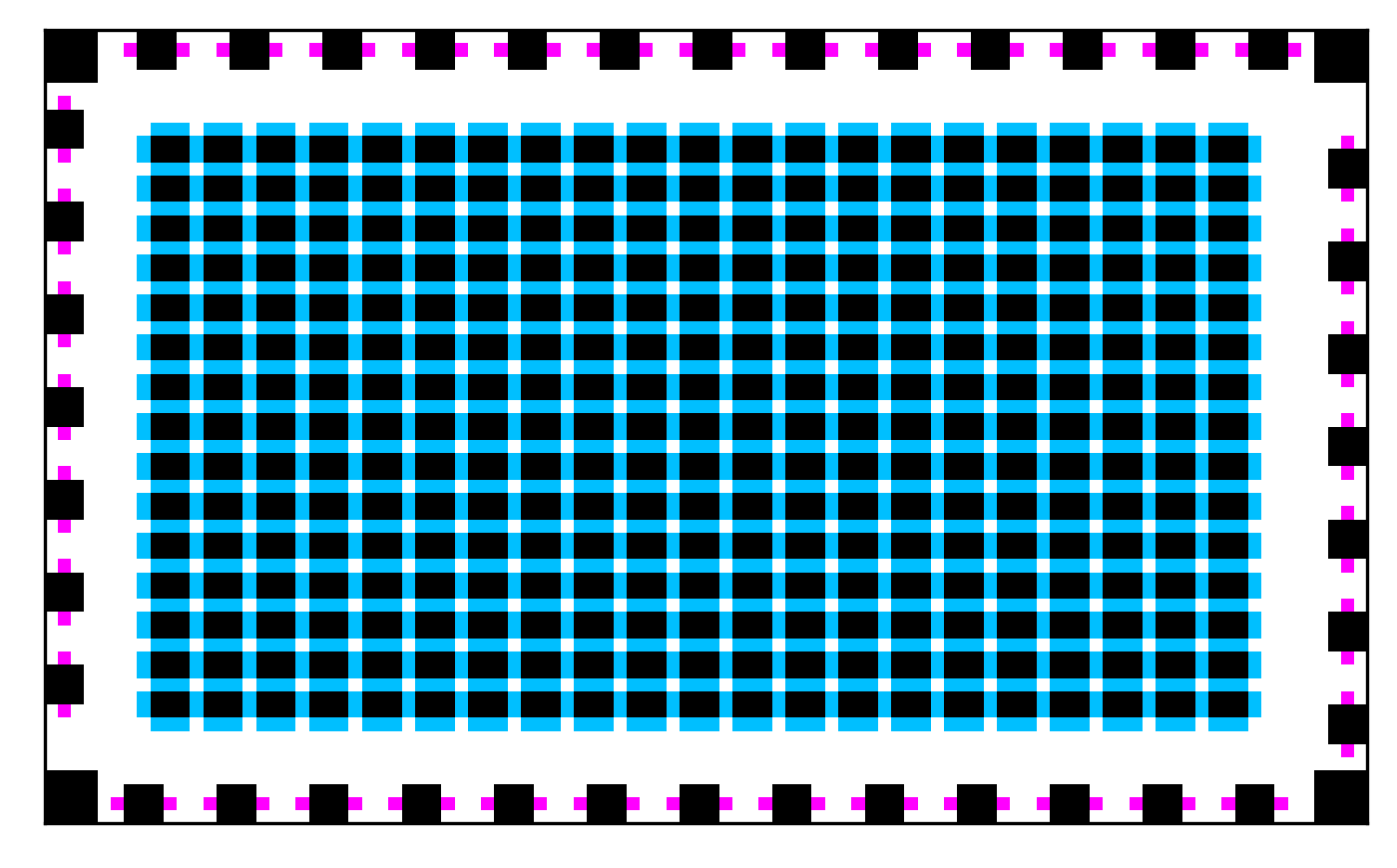}
    \includegraphics[height=0.8in]{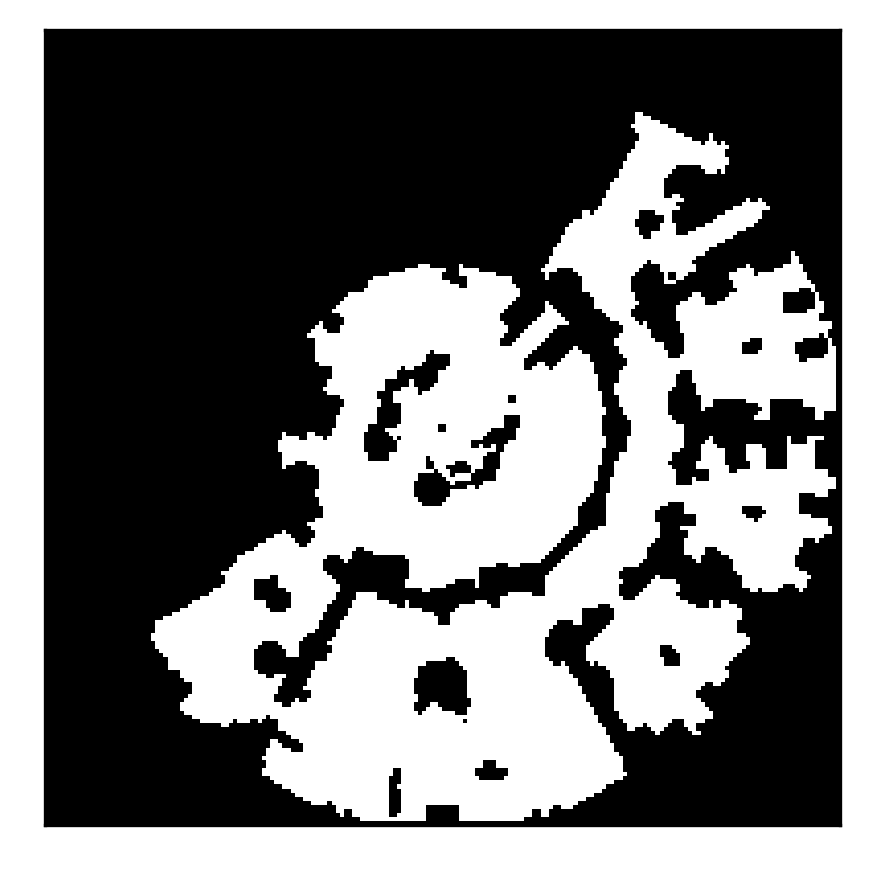}
    \includegraphics[width=0.9\linewidth]{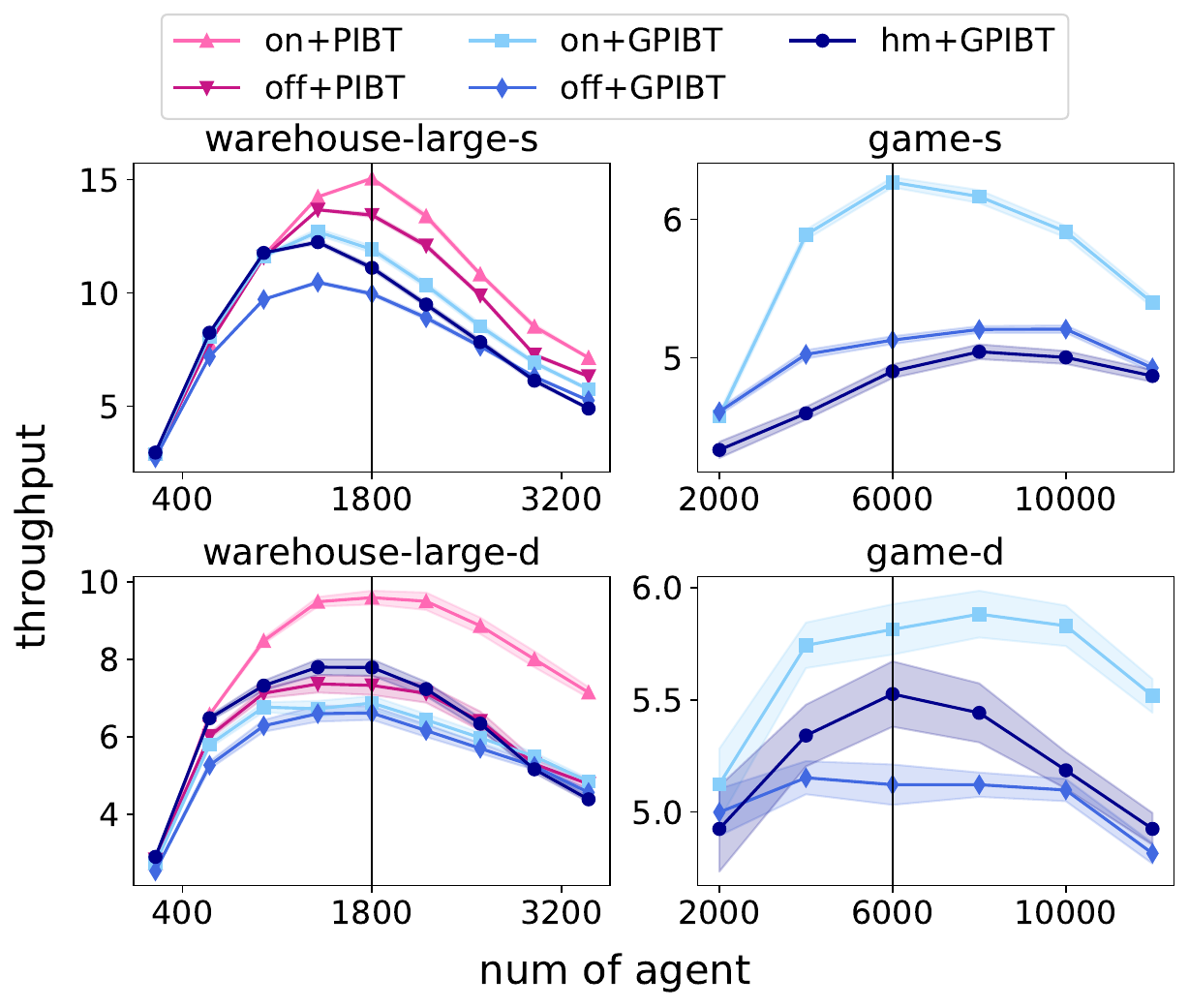}
    \caption{The visualization and the \textit{throughput-agent} curve for the \textit{warehouse-large} map and the \textit{game map}. The notation of this figure is similar to that in \Cref{fig:all}. }
    \label{fig:largemap-result}
\end{figure}
\mysubsubsection{Results on Large Maps.} We present results for larger maps, specifically the \textit{warehouse-large-60-100} map and the \textit{game-194-194} map. We design the \textit{warehouse-large} map by imitating the layout of \textit{warehouse-33-57}. The \textit{game} map is selected from the MAPF benchmark~\cite{SternSoCS19}. Visualizations of both maps are shown in \Cref{fig:largemap-result}. For the dynamic task setting, we apply a Gaussian distribution on the \textit{warehouse-large} map and a multi-modal Gaussian distribution on the \textit{game} map.

The left section of \Cref{fig:largemap-result} presents results for the \textit{warehouse-large} map. Comparing on+PIBT with off+PIBT, and on+GPIBT with off+GPIBT, it is clear that online guidance outperforms offline guidance. For the static task distribution, our optimized policy (on+GPIBT) performs better than hm+GPIBT. However, under dynamic task distribution, hm+GPIBT outperforms on+GPIBT on this map.

As the map size increases, the simulation time also increases. We do not optimize guidance for PIBT in the \textit{game} map due to computational limitations. However, GPIBT-based algorithms leverage guided paths to reduce heuristic update effort, making them adaptable to larger maps. The right section of \Cref{fig:largemap-result} shows results where on+GPIBT outperforms both off+GPIBT and hm+GPIBT.

The runtime for all algorithms is presented in \Cref{tab:runtime} in \Cref{appendix:runtime}, and the CMA-ES hyperparameters are detailed in \Cref{tab:hyperparams-large} in \Cref{appendix:exp:hyper}.


\mysubsubsection{Additional Results. } We present the ablation results for the guidance update interval $m$ in the on+PIBT algorithm in \Cref{appendix:result:update}. We discuss the deadlock issue in \Cref{appendix:deadlock}.

\subsection{Guidance Policy Visualization}
\label{sec:visualization}

In this section, we expand on earlier hypotheses and explain some results mentioned before by visualizing the guidance graph given by on+PIBT. 

\mysubsubsection{Online guidance can capture congestion locations.}
The online guidance policies outperform the offline guidance graphs because they can get more real-time information from the downstream algorithms. For instance, the online policy can capture different congestion locations and alleviate them.
\begin{figure}[!t]
    \centering
    \includegraphics[width=0.25\linewidth]{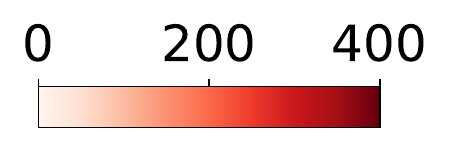}
    \includegraphics[width=0.5\linewidth]{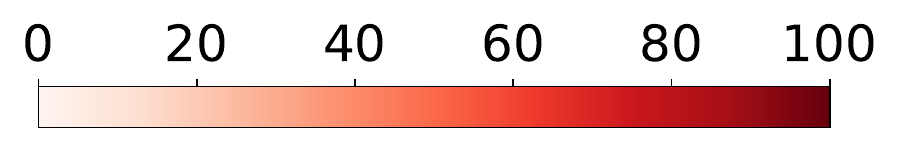}
    \subfigure[task]{
    \includegraphics[width=0.25\linewidth]{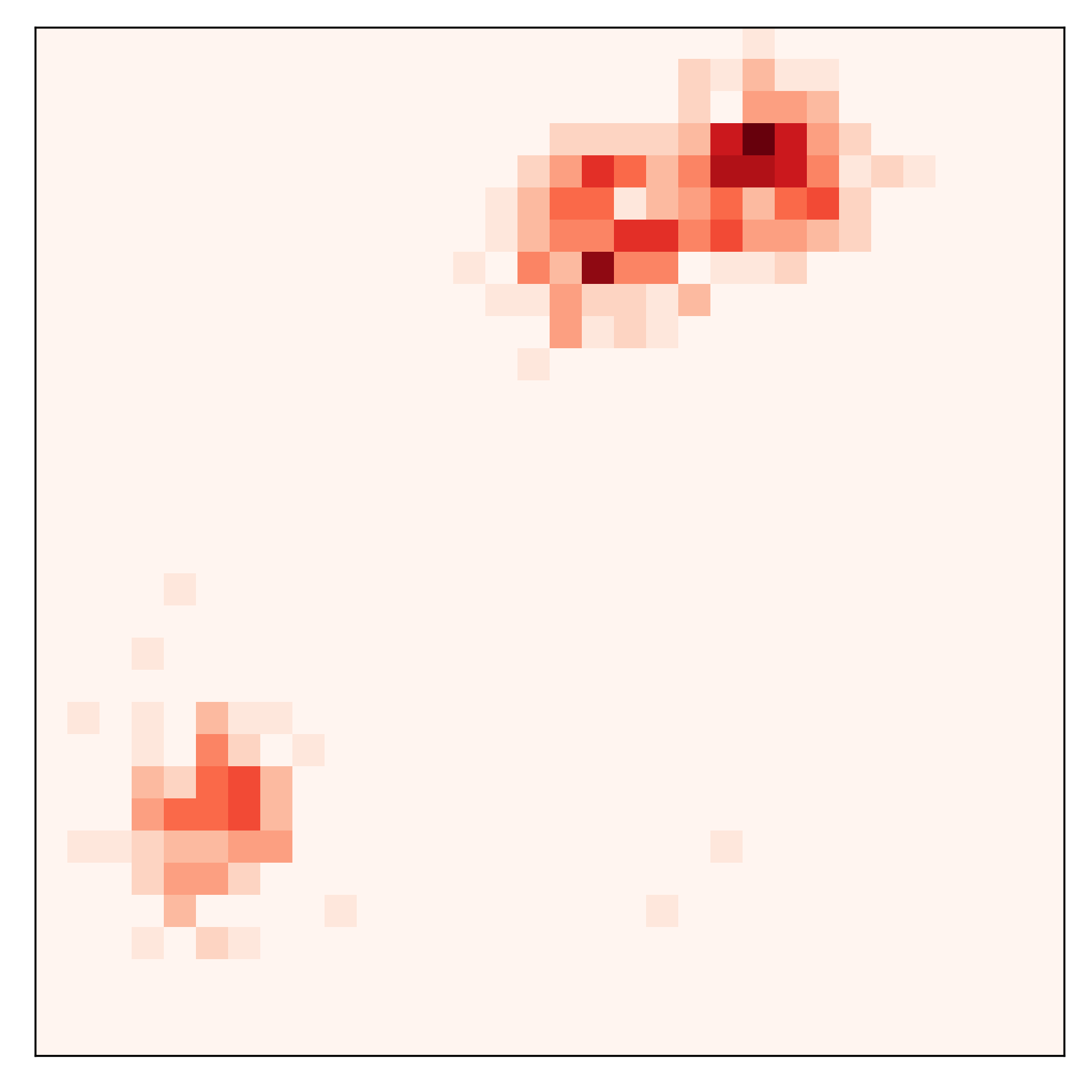}\label{fig:vis-on-off-gg:task}
    }
    \subfigure[on+PIBT]{
    \includegraphics[width=0.25\linewidth]{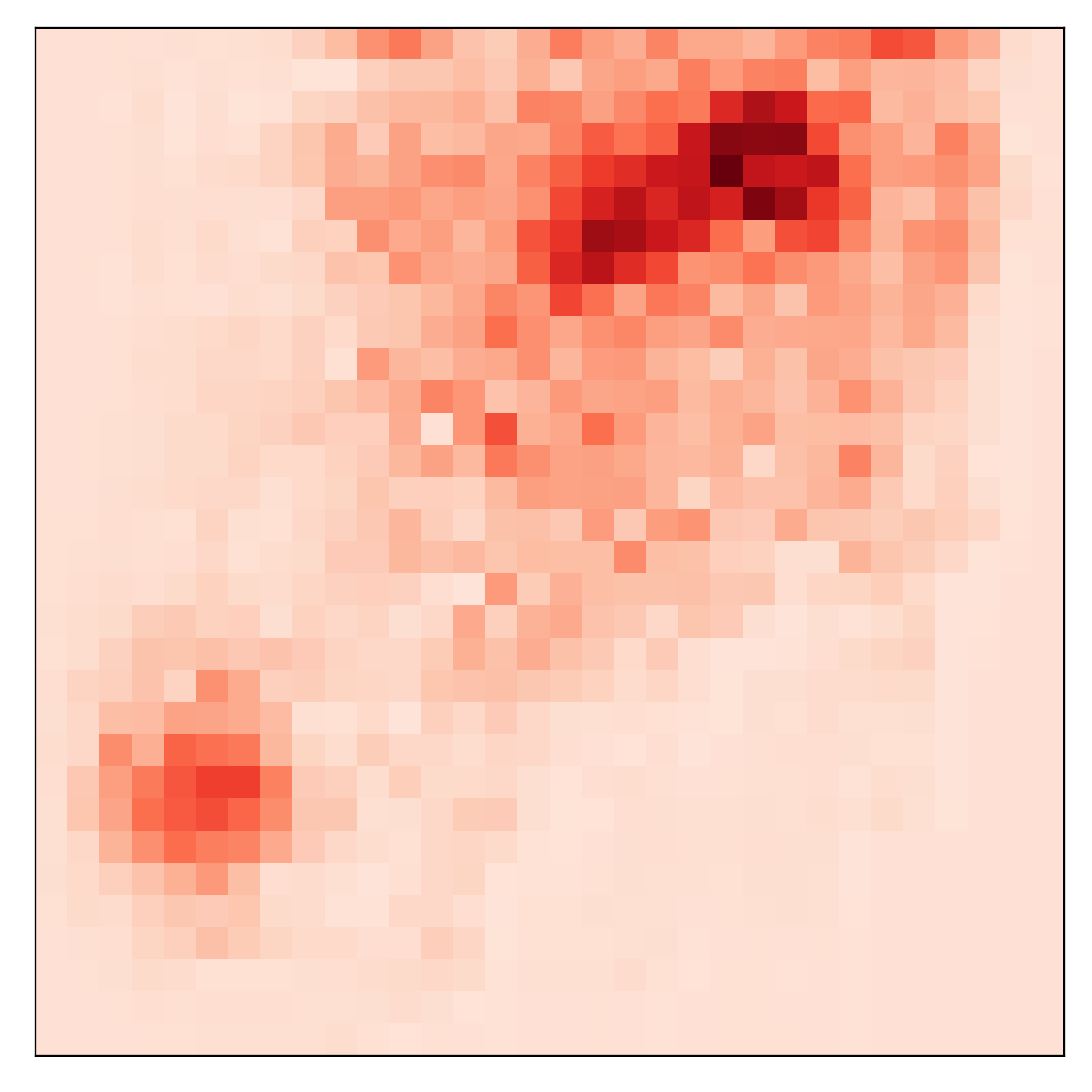}\label{fig:vis-on-off-gg:on}
    }
    \subfigure[off+PIBT]{
    \includegraphics[width=0.25\linewidth]{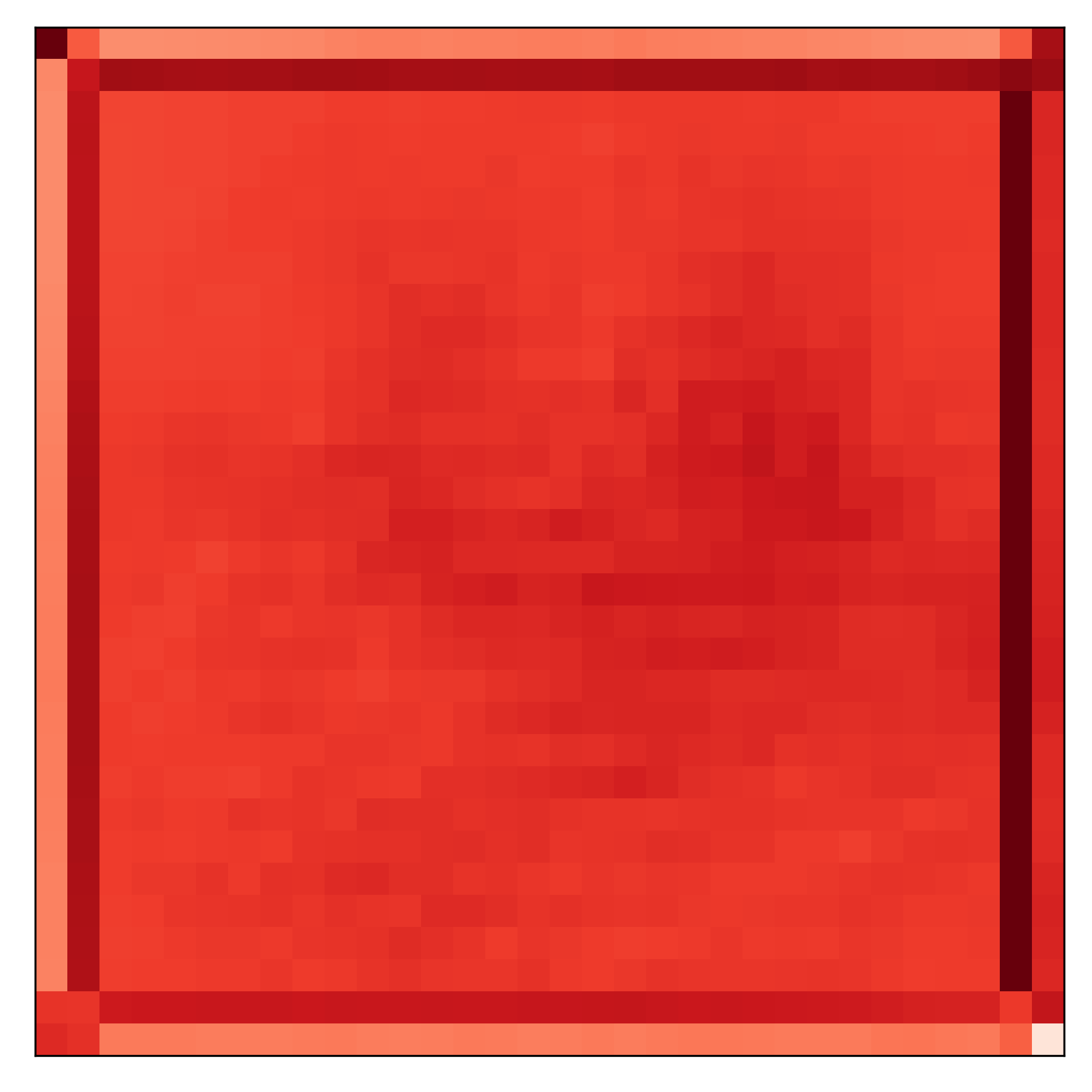}\label{fig:vis-on-off-gg:off}
    }
    \caption{Online and offline guidance given task distribution on the \textit{empty} map. Darker colors indicate more goals and higher wait costs.}
    \label{fig:vis-on-off-gg}
\end{figure}

\Cref{fig:vis-on-off-gg:on,fig:vis-on-off-gg:off} show the online and offline wait costs given the task distribution shown in \Cref{fig:vis-on-off-gg:task} on the \textit{empty} map with the dynamic task distribution. The tasks are sampled from a 3-mode Gaussian distribution. 
We observe that the wait costs given by our guidance policy align with the task distribution. Specifically, the areas with more tasks are expected to be more congested. Therefore, our guidance policy generates higher wait costs to encourage the agents to move away from such areas, alleviating congestion.
The wait costs given by the offline guidance graph, however, have no correlation with the task distribution at all.

\begin{figure}[!h]
    \centering
    \includegraphics[width=0.8\linewidth]{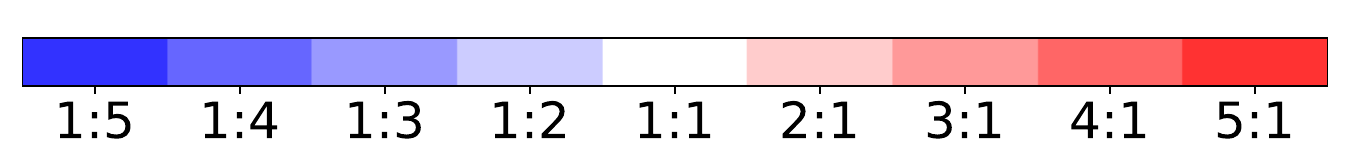}
    \subfigure[sortation-on]{
    \includegraphics[width=0.4\linewidth]{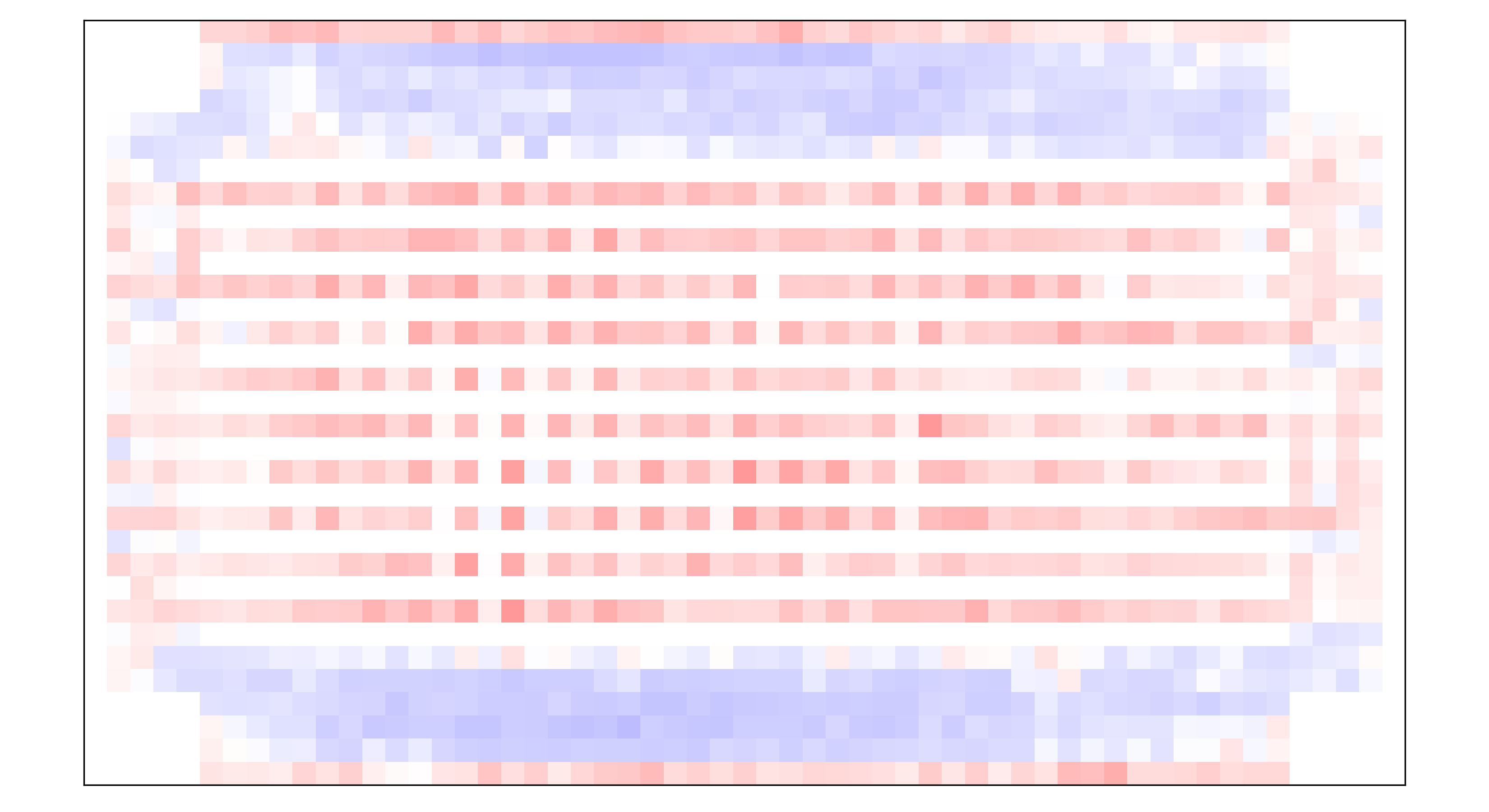}
    }
    \subfigure[warehouse-on]{
    \includegraphics[width=0.4\linewidth]{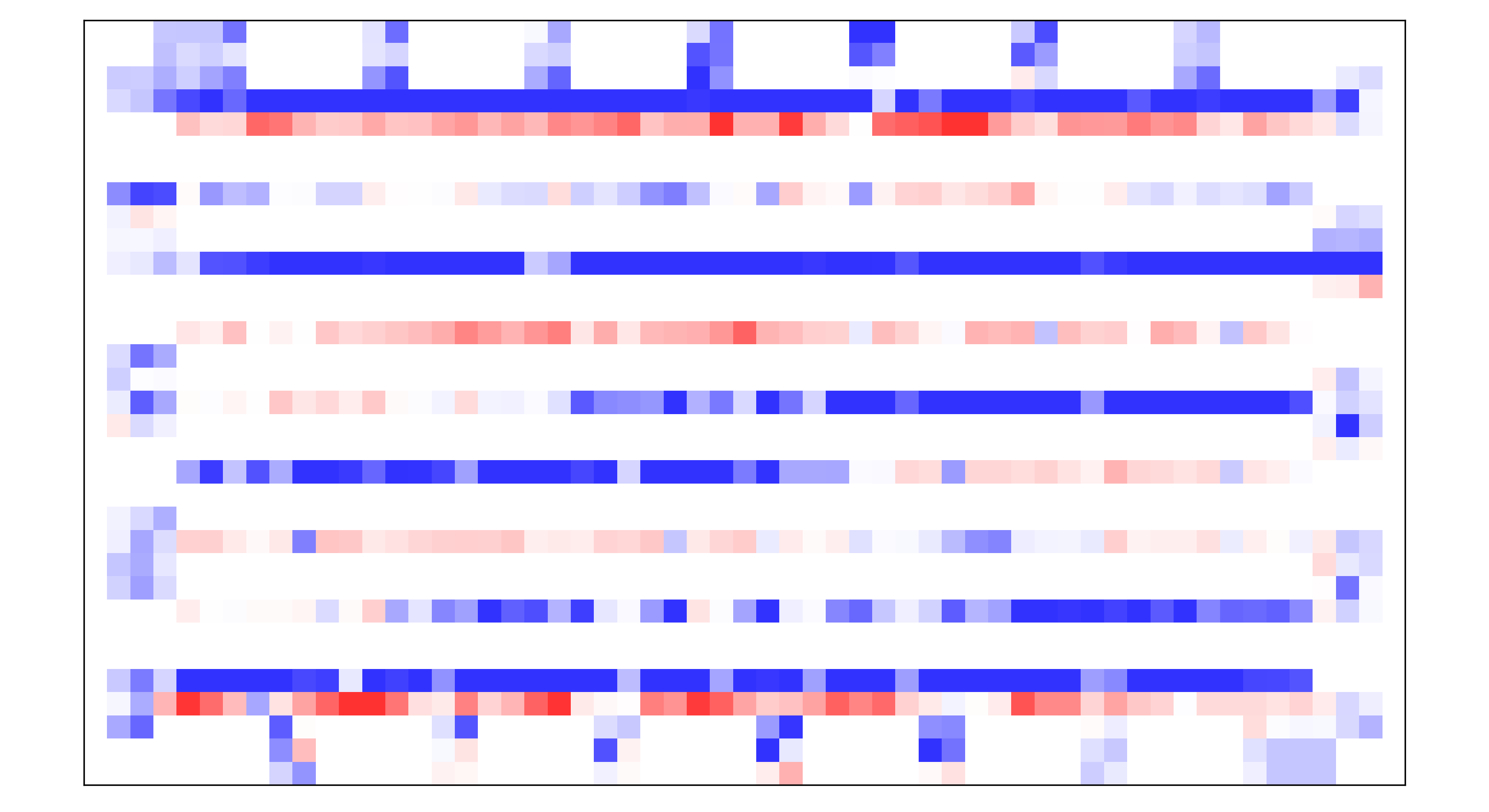}
    }\\
    
    \subfigure[sortation-off]{
    \includegraphics[width=0.4\linewidth]{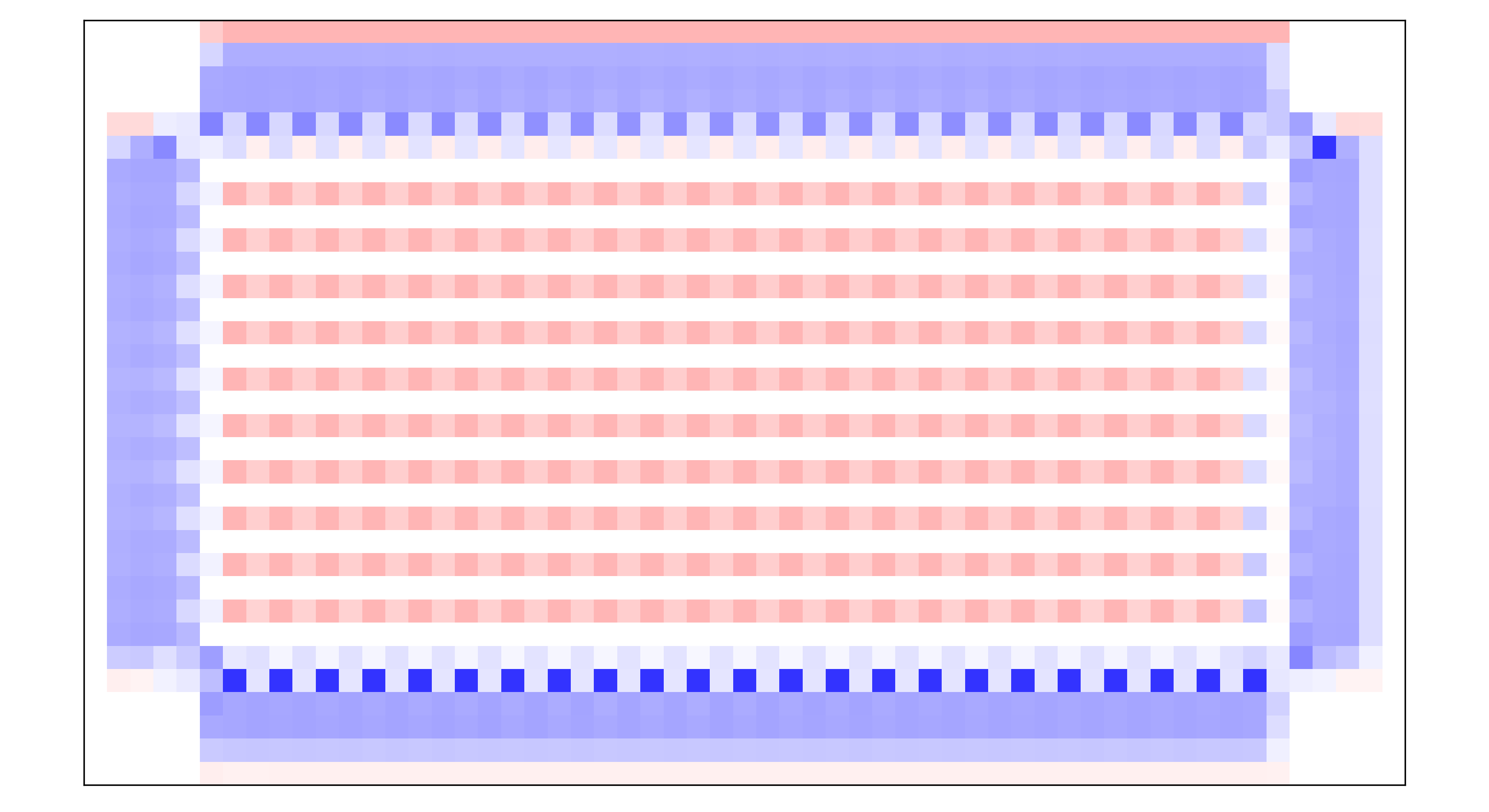}
    }
    \subfigure[warehouse-off]{
    \includegraphics[width=0.4\linewidth]{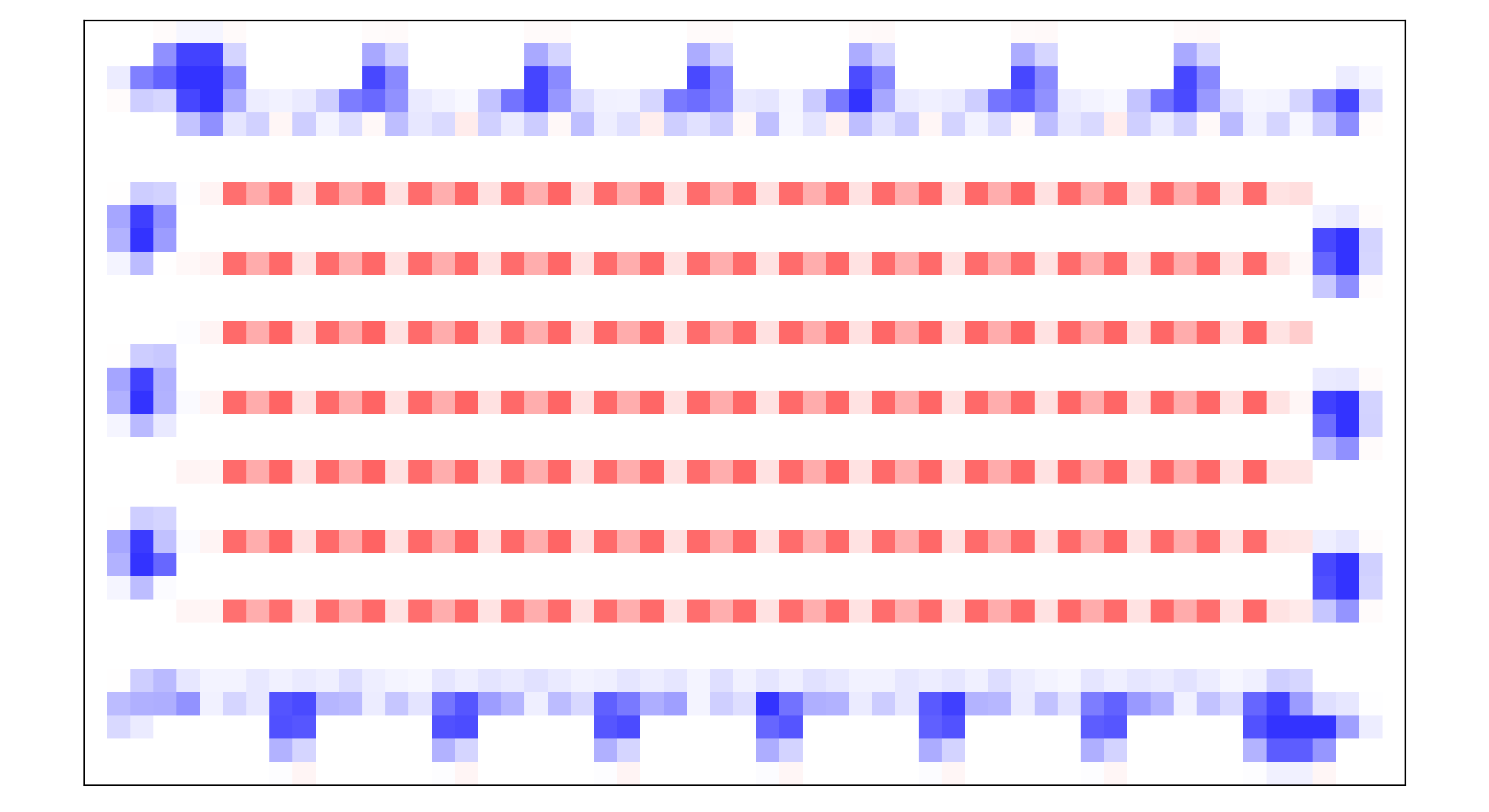}
    }
    \caption{Visualization of the ratio of left to right edge weights in the guidance graph for \textit{sortation} and \textit{warehouse} maps, where ``on'' for on+PIBT and ``off'' for off+PIBT. Red cells indicate a higher ratio of left to right, while blue cells indicate a higher ratio of right to left.}
    \label{fig:sort-warehouse}
\end{figure}
\mysubsubsection{The structure of the map affects the improvement ratio of online guidance policy.}
Although the \textit{sortation} and \textit{warehouse} maps look alike, they have different properties and thus lead to different performances in throughput. In \textit{warehouse}, the white space between workstations (pink) and endpoints (blue) is more narrow, and the
corridors among endpoints are longer. The former makes detours in the white space less useful, and the latter decreases the number of path choices when agents are in the center of the map. Therefore, it is important for the online guidance policy to lead agents to choose an appropriate corridor and direction when entering the central area of the \textit{warehouse} map to avoid congestion. Under the static task distribution, we can clearly see the alternative preferred direction on \textit{warehouse} map in the generated online guidance, but can hardly capture it in \textit{sortation} map. (See \Cref{fig:sort-warehouse}). On the other hand, in both maps, offline guidance graphs direct agents in one direction in the central area and in another direction in the surrounding area.  
Therefore, we can see why, under the static task distribution setting, online guidance leads to dramatic improvement on the \textit{warehouse} map but not on the \textit{sortation} map. 



\section{Conclusion}
Lifelong MAPF is a challenging problem of immense practical importance. Once we have a large number of agents, all practical approaches to the problem must overcome congestion problems. In this paper, we show how we can use dynamically generated guidance graphs that learn from traffic patterns to generate guidance, alleviating traffic congestion and improving throughput. We show that our proposed online guidance policy improve throughput in two state-of-the-art LMAPF algorithms across 4 different maps. Our work reveals several future directions. First, future works can explore incorporating the guidance policy in other LMAPF algorithms such as RHCR~\cite{Li2020LifelongMP} and Learn to Follow~\cite{learntofollow2024}. Second, our work leverages CMA-ES, requiring a large number of evaluations in the LMAPF simulators to optimize the guidance policy. Future works can explore surrogate-assisted optimizers~\cite{Zhang2021DeepSA,Bhatt2022DeepSA,kent2020bop} to improve the sample efficiency of the optimization.

\section*{Acknowledgments}
Hongzhi Zang performed her research during her visit to Carnegie Mellon University.
This work used Bridge-2 at Pittsburgh Supercomputing Center (PSC) through allocation CIS220115 from the Advanced Cyberinfrastructure Coordination Ecosystem: Services \& Support (ACCESS) program, which is supported by National Science Foundation grants \#2138259, \#2138286, \#2138307, \#2137603, and \#2138296.

\bibliography{aaai25}

\clearpage

\appendix

\clearpage

\begin{table}[!t]
\centering
\resizebox{1\linewidth}{!}{
\begin{tabular}{c|c|rrr}
\toprule
map-[dist type] & algorithm           & $N_{eval}$ & $b$  & $N_e$ \\\midrule
\textit{sortation}-s    & on+PIBT        & 50,000    & 50                 & 3        \\
                & off+PIBT       & 50,000    & 50                 & 3        \\
                & on+GPIBT       & 10,000    & 100                 & 2        \\
                & {[}p-on]+GPIBT & 10,000    & 100                 & 2        \\
                & off+GPIBT      & 10,000    & 100                 & 2        \\\midrule
\textit{sortation}-d    & on+PIBT        & 12,500    & 50                 & 3        \\
                & off+PIBT       & 12,500    & 50                 & 3        \\
                & on+GPIBT       & 50,000    & 100                 & 5        \\
                & {[}p-on]+GPIBT & 50,000    & 100                 & 5        \\
                & off+GPIBT      & 50,000    & 100                 & 5        \\\midrule
\textit{warehouse}-s    & on+PIBT        & 25,000    & 50                 & 3        \\
                & off+PIBT       & 25,000    & 50               & 3        \\
                & on+GPIBT       & 20,000    & 100              & 2        \\
                & {[}p-on]+GPIBT & 20,000    & 100                 & 2        \\
                & off+GPIBT      & 20,000    & 100                & 2        \\\midrule
\textit{warehouse}-d    & on+PIBT        & 25,000    & 50              & 3        \\
                & off+PIBT       & 25,000    & 50                & 3        \\
                & on+GPIBT       & 50,000    & 100                & 2        \\
                & {[}p-on]+GPIBT & 50,000    & 100                 & 2        \\
                & off+GPIBT      & 50,000    & 100                 & 2        \\\midrule
\textit{empty}-s    & on+PIBT        & 25,000    & 50                 & 3        \\
                & off+PIBT       & 25,000    & 50                 & 3        \\
                & on+GPIBT       & 10,000    & 100                 & 2        \\
                & {[}p-on]+GPIBT & 10,000    & 100                 & 2        \\
                & off+GPIBT      & 10,000    & 100                 & 2        \\\midrule
\textit{empty}-d    & on+PIBT        & 25,000    & 50                 & 3        \\
                & off+PIBT       & 25,000    & 50                 & 3        \\
                & on+GPIBT       & 10,000    & 100                 & 2        \\
                & {[}p-on]+GPIBT & 10,000    & 100                 & 2        \\
                & off+GPIBT      & 10,000    & 100                 & 2        \\\midrule
\textit{random}-s    & on+PIBT        & 50,000    & 50                & 3        \\
                & off+PIBT       & 50,000    & 50                 & 3        \\
                & on+GPIBT       & 50,000   & 100                & 5        \\
                & {[}p-on]+GPIBT & 50,000    & 100                & 5        \\
                & off+GPIBT      & 50,000   & 100                 & 5        \\\midrule
\textit{random}-d    & on+PIBT        & 50,000    & 50                 & 3        \\
                & off+PIBT       & 50,000    & 50                & 3        \\
                & on+GPIBT       & 20,000    & 100                & 5        \\
                & {[}p-on]+GPIBT & 20,000    & 100                 & 5        \\
                & off+GPIBT      & 20,000    & 100                 & 5        \\
\bottomrule
\end{tabular}
}
\caption{CMA-ES hyperparameters. [dist type] is an abbreviation for distribution type, where ``d'' stands for dynamic and ``s'' stands for static.}
\label{tab:hyperparams}
\end{table}
\begin{table}[!t]
\centering
\resizebox{1\linewidth}{!}{
\begin{tabular}{c|c|rrr}
\toprule
map-[dist type] & algorithm           & $N_{eval}$ & $b$  & $N_e$ \\\midrule
\textit{warehouse-large}-s  & on+PIBT & 10,000  & 50    & 3 \\
                            & off+PIBT & 10,000 & 50    & 3 \\
                            & on+GPIBT &  10,000  & 100  & 2  \\
                            & {[}p-on]+GPIBT & 10,000 & 100  & 2\\
                            & off+GPIBT &  10,000 & 100 & 2\\\midrule
\textit{warehouse-large}-d  & on+PIBT & 10,000  & 50    & 3 \\
                            & off+PIBT & 10,000 & 50    & 3 \\
                            & on+GPIBT & 20,000   & 100  & 5  \\
                            & {[}p-on]+GPIBT & 20,000 & 100  & 5 \\
                            & off+GPIBT & 20,000 & 100  & 5\\\midrule
\textit{game}-s & on+GPIBT & 10,000 & 100  &   2\\
                & off+GPIBT & 10,000    & 100  & 2\\\midrule
\textit{game}-d & on+GPIBT & 10,000 & 100  &   2\\
                & off+GPIBT & 10,000    & 100  & 2\\
\bottomrule
\end{tabular}
}
\caption{CMA-ES hyperparameters for the \textit{warehouse-large} and \textit{game} map.}
\label{tab:hyperparams-large}
\end{table}

\section{Detailed Experiments Setups}

In this section, we present the detailed setups of the experiments presented in \Cref{sec:experiment}.

\subsection{Hyperparameters} \label{appendix:exp:hyper}

\begin{table}[!t]
\centering
\resizebox{1\linewidth}{!}{
\begin{tabular}{c|cccc}
\toprule
      & \textit{sortation} & \textit{warehouse} & \textit{empty} & \textit{random} \\
\midrule
$\mu$ selection & \textit{ep} & \textit{ep} & \textit{ap} & \textit{ap} \\
$\sigma$ & 1.0       & 1.0       & 0.5   & 0.5    \\
K     & 1         & 1         & 3     & 3      \\
$N_d$ & 200       & 200       & 200   & 200    \\
\bottomrule
\end{tabular}
}
\caption{Dynamic task distribution hyperparameters. \textit{ep} stands for endpoints. \textit{ap} stands for any locations on the whole map. $\mu$ and $\sigma$ denote the mean and standard deviation of Gaussian distribution, $K$ denotes the $K$-mode Gaussian, and $N_d$ denotes the task distribution change interval.}
\label{tab:dist-hyper}
\end{table}

\mysubsubsection{CMA-ES.}
\Cref{tab:hyperparams,tab:hyperparams-large} show the hyperparameters in CMA-ES. $b$ stands for batch size, $N_{e}$ stands for the number of simulations used to evaluate one solution sampled from the CMA-ES Gaussian distribution, and $N_{eval}$ stands for the total number of evaluations during the optimization. 
We select various values of $b$ to accommodate the number of parameters in the guidance policies for PIBT and GPIBT. Additionally, we choose different values of $N_e$ to mitigate the variance of the CMA-ES objective function in each evaluation. The rationale behind using different $N_{eval}$ values is that guidance policies in different settings have varying convergence requirements in terms of the total number of evaluations needed.
While the hyperparameters vary, we use the same hyperparameters for the settings that we make comparisons. 

\mysubsubsection{Task setup.}
As mentioned in \Cref{sec:exp-overview}, we apply Gaussian and multi-modal Gaussian as the dynamic task distributions on different maps. \Cref{tab:dist-hyper} shows the detailed hyperparameters. To sample the tasks, we regard the map as a 2D space. The Gaussian center is selected from a subset of all locations on the map, as indicated in the table. 
After $N_d$ timesteps, the Gaussian centers are randomly sampled again, and agents' new tasks are sampled from the updated (multi-modal) Gaussian distribution.

\subsection{On+PIBT Guidance Policy} \label{appendix:exp:on-pibt-gp}
The guidance policy for on+PIBT is a 3-layer Convolutional Neural Network (CNN). The kernel sizes on each layer are 3, 1, and 1, respectively. We choose ReLU and BatchNorm2d to serve as the non-linearity layers. The observation of the guidance policy includes past traffic and current tasks. Concretely, the past traffic is the edge usage of all edges in $E_g$ in previous $m$ steps. It is encoded in shape $(5, h, w)$, where $5$ denotes the outgoing edges in 5 directions (right, up, left, down, self) of each vertex, and $(h, w)$ denotes the height and width of the map. The current tasks encode the current goal locations for all agents with the shape of $(h, w)$. For location $(x, y)$, the value is the number of goals on $(x, y)$. Each channel of the observation is normalized in $(0, 1)$ to stabilize the model output. The output of the guidance policy is the guidance graph edge weights of shape $(5, h, w)$. The total number of parameters is 3,119. 

We use a small CNN  with small kernel sizes, not only because the model's observation is a 3D tensor, but also to ensure the number of parameters is not too large so that it is easier for CMA-ES to optimize. 

\subsection{On+GPIBT Guidance Policy}\label{appendix:exp:on-gpibt-gp}
\mysubsubsection{Guidance policy implementation. }
As mentioned in \Cref{sec:approach}, we call the model to update the guidance graph before an agent plans its guide path. In implementation, we actually use a \textit{lazy} mechanism. We calculate the edge weights in the guidance graph only for those edges that connect to the expanded neighbors of the current locations during the search for the shortest guide paths on the graph. This lazy mechanism avoids computing unnecessary edge weights, thereby reducing the overhead of calling the guidance policy during agents' guide path generation. 
\Cref{algo:search-gp} shows the pseudo-code of the search process for the guide path of each agent. Overall, it is an A* search. We employ $h(x)\equiv 1$ as the admissible heuristic function, given that all edge weights in the guidance graph are at least 1. \Cref{algo:search:init1} and \Cref{algo:search:init2} initialize the OPEN list with the start location $s$ and the CLOSE list with an empty list. \Cref{algo:search:while} to \Cref{algo:search:end-while} is the main \textit{while} loop of the A* search. \Cref{algo:search:pop1} pops the node with least $f$ values (the sum of $g$ and $h$). \Cref{algo:search:pop2} then push the popped node into the close list. These are the preparation steps before expanding the \textit{curr} node. \Cref{algo:search:check-goal} checks whether the expanded node is the goal location. We then compute the model observation (\Cref{algo:search:win-obs}) and use the guidance policy to compute the edge weights on the guidance graph (\Cref{algo:search:model}). We explain the windowed observation in the next subsection. Using the weights, we continue to expand the current node. \Cref{algo:search:expand} to \Cref{algo:search:end-for} is the \textit{for} loop for the expanding process, the same as the typical A* search.  

\begin{algorithm}[!t]
\caption{Search for guide paths by using guidance policy}
\label{algo:search-gp}
    \begin{algorithmic}[1]
        \STATE \textbf{Input} start location $s$; goal location $goal$; online policy $\pi_\theta$; current all agents guide-path edge usage $U_e$
        \STATE \textbf{Notation} OPEN, open list in A* search, which is a priority heap; CLOSE, the closed list in the search, ensures that no node is expanded more than once; $h(x)$, any admissible heuristic function for location $x$.
        \STATE root $\gets \text{search-node(loc=s, g=0, h=h(s), parent=none)}$ \label{algo:search:init1}
        \STATE OPEN $\gets \{\text{root}\}$; CLOSE $\gets\{\}$ \textit{// initialize open list}\label{algo:search:init2}
        \WHILE{OPEN $\neq \phi$}
        \label{algo:search:while}
            \STATE curr $\gets$ OPEN.pop() \label{algo:search:pop1}
            \STATE CLOSE.push(curr)\label{algo:search:pop2}
            \IF{curr.loc$=goal$}\label{algo:search:check-goal}
                \STATE break
            \ENDIF
            \STATE $obs\gets$ Windowed-Obs(curr.loc, win\_size, $U_e$) \label{algo:search:win-obs}
            \STATE $cost \gets \pi_\theta(obs) + 1$. \textit{// {$cost\in\mathbb{R}^4$}} \label{algo:search:model}
            \FOR{$nid, n$ in $enumerate(\text{curr.loc.neighbors})$} \label{algo:search:expand}
                \STATE $cost_n = cost[nid]$
                \STATE next $\gets \text{search-node}(\text{loc=}n, \text{g=}\text{curr.g+}cost_n, h=h(n), \text{parent=curr})$
                \IF{$n$ is not equal to the location of any nodes in OPEN }
                    \STATE OPEN.push(next)
                \ELSE
                    \STATE \textit{// $\exists\text{existing}\in\text{OPEN}$ s.t. existing.loc$=$next}
                    \IF{next.g $<$ existing.g}
                    \STATE assert existing $\notin$ CLOSE
                    \STATE OPEN.update(existing, next) \textit{// use ``next'' to replace the ``existing'' node in the open list. }
                    \ENDIF
                \ENDIF
            \ENDFOR\label{algo:search:end-for}
        \ENDWHILE\label{algo:search:end-while}
    \end{algorithmic}
\end{algorithm}

\mysubsubsection{Guidance policy model parameters. }
We use $s_{win}$ to denote the window size of the observation. The observation of the guidance policy is the windowed guide-paths edge usage, encoded in the shape of (4, $s_{win}$, $s_{win}$). The window center is located at the current expanding node. 4 channels stand for right, up, left, and down guide-paths edge usage of other agents. We choose $s_{win}$=5. The output of the model is the weights of the guidance graph on 4 outgoing edges of the current expanding node. 

We use a simple quadratic function as guidance policy. Every term in the windowed guide-path edge usage is associated with an optimizable linear variable. We further include quadratic variables on \textit{contra-flow} of the windowed guide-path edge usage. That is, for each pair of adjacent vertex $u$ and $v$ included in the guide-path edge usage, we consider the product term of guide-path edge usage on $(u, v)$ and $(v, u)$. 
Such a design is inspired by \cite{ChenAAAI24}, and it ensures that the handcrafted function in hm+GPIBT can be represented by our network architecture.
The total number of parameters for on+GPIBT online guidance policy model is 560. 

\subsection{Compute Resource} \label{appendix:exp:compute}
We run our experiments on 4 machines: (1) a local machine with a 64-core AMD Ryzen Threadripper 3990X CPU, 192 GB of RAM, and an Nvidia RTX 3090Ti GPU, (2) a local machine with a 64-core AMD Threadripper 7980X CPU, 128 GB of RAM, and an Nvidia RTX 1080Ti GPU, (3) a Nectar Cloud instance with a 32-core AMD EPYC-Rome Processor and 64GB RAM, and (4) a high-performing cluster with numerous 64-core AMD EPYC 7742 CPUs, each with 256GB of RAM. The CPU runtime is measured on the machine (1).

\begin{table*}[htbp]
\centering
\begin{tabular}{r|rr|rrrr}
\toprule
map-type     &  on+PIBT       & off+PIBT      & on+GPIBT      & {[}p-on]+GPIBT & off+GPIBT    & hm+GPIBT      \\
\midrule
\textit{sortation}-s    &  27.912$\pm$5.504  & \textbf{11.284$\pm$0.299} & 16.752$\pm$0.214  & 5.735$\pm$0.698   & \textbf{2.943$\pm$0.078} & 3.326$\pm$0.131  \\
\textit{sortation}-d    & 24.954$\pm$4.687  & \textbf{8.016$\pm$0.765}   & 9.504$\pm$0.578  & 5.117$\pm$0.834   & \textbf{2.176$\pm$0.075} & 2.605$\pm$0.104  \\
\textit{warehouse}-s    & 15.515$\pm$2.682 & \textbf{5.883$\pm$0.306}  & 7.902$\pm$0.265  & 3.826$\pm$0.692   & \textbf{1.440$\pm$0.029} & 1.475$\pm$0.036  \\
\textit{warehouse}-d    & 14.061$\pm$2.238 & \textbf{4.395$\pm$0.217}  & 4.074$\pm$0.269  & 3.618$\pm$0.715   & \textbf{1.203$\pm$0.047} & 1.292$\pm$0.0483 \\
\textit{empty}-s    & 11.416$\pm$1.685 & \textbf{6.405$\pm$0.196}  & 10.322$\pm$0.317 & 3.665$\pm$0.603 & \textbf{1.787$\pm$0.066} & 2.026$\pm$0.092  \\
\textit{empty}-d    & 9.944$\pm$1.408  & \textbf{4.439$\pm$0.306}  & 8.482$\pm$0.548  & 3.149$\pm$0.636    & \textbf{1.200$\pm$0.056} & 1.267$\pm$0.051  \\
\textit{random}-s    & 8.524$\pm$1.182  & \textbf{4.804$\pm$0.332}  & 5.912$\pm$1.132  & 3.173$\pm$0.696   & \textbf{1.279$\pm$0.063} & 1.348$\pm$0.089  \\
\textit{random}-d    & 7.706$\pm$0.965  & \textbf{3.638$\pm$0.442}  & 3.530$\pm$0.323  & 2.857$\pm$0.657   & \textbf{0.951$\pm$0.046} & 0.953$\pm$0.045 \\
\textit{warehouse-large}-s & 115.342$\pm$30.211	& 25.496$\pm$3.747 & 20.692$\pm$0.426 & 11.200$\pm$0.818 &	\textbf{7.313$\pm$0.121}	& 8.678$\pm$0.213 \\
\textit{warehouse-large}-d & 113.408$\pm$30.592 & 18.266$\pm$2.004&	13.792$\pm$0.743&	9.467$\pm$0.789 & \textbf{5.845$\pm$0.292} &	7.640$\pm$0.745 \\
\textit{game}-u & - & - & 63.409$\pm$1.373 & 38.252$\pm$0.770 & 30.664$\pm$0.531 & \textbf{29.015$\pm$0.849} \\
\textit{game}-d & - & - & 63.502$\pm$2.403 & 37.247$\pm$0.786 &	30.277$\pm$0.553 & \textbf{28.960$\pm$0.991}\\
\bottomrule
\end{tabular}
\caption{CPU runtime (in seconds) table for all algorithms for 1,000 timesteps. Type $s$ and $d$ stand for static and dynamic task distribution. The numbers of agents are 800, 600, 400, 400, 1800, and 6000 for the \textit{sortation}, \textit{warehouse}, \textit{empty}, \textit{random}, \textit{warehouse-large}, and \textit{game} map.}
\label{tab:runtime}
\end{table*}
\subsection{Relevant Software Library} \label{appendix:exp:software}
We follow the previous work~\cite{zhang2024ggo} and implement CMA-ES using Pyribs~\cite{pyribs}. We use the open-source Guided-PIBT implementation of \citet{ChenAAAI24}, and PIBT implementation of \citet{Jiang2024Competition}.
\section{Additional Experimental Results}
In this section, we show the following additional experiment results: 
(1) runtime of different algorithms, (2) ablation results on the guidance update interval $m$, and (3) visualization for deadlock mitigation.
\subsection{Runtime}
\label{appendix:runtime}
\Cref{tab:runtime} shows numerical results for runtime mentioned in \Cref{subsec:exp-result}. The runtime of on+PIBT is 2-3 times slower than off+PIBT due to the need to recompute the heuristics. For GPIBT-based methods, the runtime of off+GPIBT and hm+GPIBT is similar, as they require minimal effort to compute the edge weights of the guidance graph when planning the guide paths. [p-on]+GPIBT does not compute edge weights during the guide-path search, but it does compute observations when updating the guidance graph. on+GPIBT is up to 7 times slower than hm+GPIBT and off+GPIBT because it updates the guidance graph every time an agent replans its guide path. 
on+GPIBT is also slower than [p-on]+GPIBT because, in [p-on]+GPIBT, all agents share the same guidance graph, which ignores the guide-path edge usage, whereas on+GPIBT does not. As a result, on+GPIBT updates the guidance graph more frequently, and thus more time-consuming. 
While on+PIBT and on+GPIBT incur higher runtimes, they achieve higher throughput.

\begin{table}[htbp]
    \centering
    \resizebox{1\linewidth}{!}{
    \begin{tabular}{c|rrrr}
    \toprule
         algo   &  on   & on(eval)  & off   & hm\\\midrule
         static &42.145$\pm$0.562 & 43.093$\pm$0.362 &\textbf{7.828$\pm$0.194} & 8.383$\pm$0.111\\
         dynamic & 38.553$\pm$1.528 & 41.290$\pm$0.980 &\textbf{6.680$\pm$0.364} & 8.961$\pm$0.321\\
        \bottomrule
    \end{tabular}
    }
    \caption{CPU runtime (in seconds) of LNS-based algorithms with 600 agents for 1000 steps. 
    \textit{Static} and \textit{dynamic} indicate the task distribution. ``on'',  ``on(eval)'', ``off'', and ``hm'' stand for on+GPIBT+LNS, on+GPIBT+LNS(eval), off+GPIBT+LNS, and hm+GPIBT+LNS, respectively. }
    \label{tab:lns-runtime}
\end{table}
\Cref{tab:lns-runtime} presents numerical results for the runtime of LNS-related algorithms mentioned in \Cref{subsec:exp-result}. 

\subsection{Explanation for hm+GPIBT's Near Optimality}
\label{appendix:hm}

\Cref{fig:all} and \Cref{fig:largemap-result} show that, under the \textit{sortation}, \textit{empty}, \textit{random}, and \textit{warehouse-large} maps, on+GPIBT and hm+GPIBT exhibit similar performance, with hm+GPIBT performing slightly better in some cases. To further investigate, we constrain the number of parameters in the on+GPIBT guidance policy.

To compute the weights $\omegaV\in \mathbb{R}_{>0}^{|E_g|}$ on guidance graph $G_g(V_g, E_g, \omegaV)$ for graph $G(V, E)$, hm+GPIBT uses the following equation. Given all guide-path edge usage $U:E\to\mathbb{R}_{\geq0}$, for any $(u, v)\in E$, 
\begin{align*}
    \omegaV_{(u, v)} = U(u, v) U(v, u)+U(v, u) + \frac{1}{2}\sum_{u'\in N(v)} U(v, u'),
\end{align*}
where $N(v)$ represents the neighbor set of $v$ in graph $G$. To constrain the parameter space, the function for on+GPIBT is defined as follows for a neighbor $u_i\in N(v)$
\begin{align*}
    \omegaV_{(u_i, v)} &= \sum_{u_j\in N(v)}p_{i, j, 1}U(u_j, v) U(v, u_j) \\
    &+\sum_{u_j\in N(v)}p_{i, j, 2}U(v, u_j) \\
    &+ \sum_{u_j\in N(v)}p_{i, j, 3}U(v, u_j),
\end{align*}
where $p_{i, j, k}$ are optimizable parameters. 

\begin{figure}
    \centering
    \includegraphics[width=\linewidth]{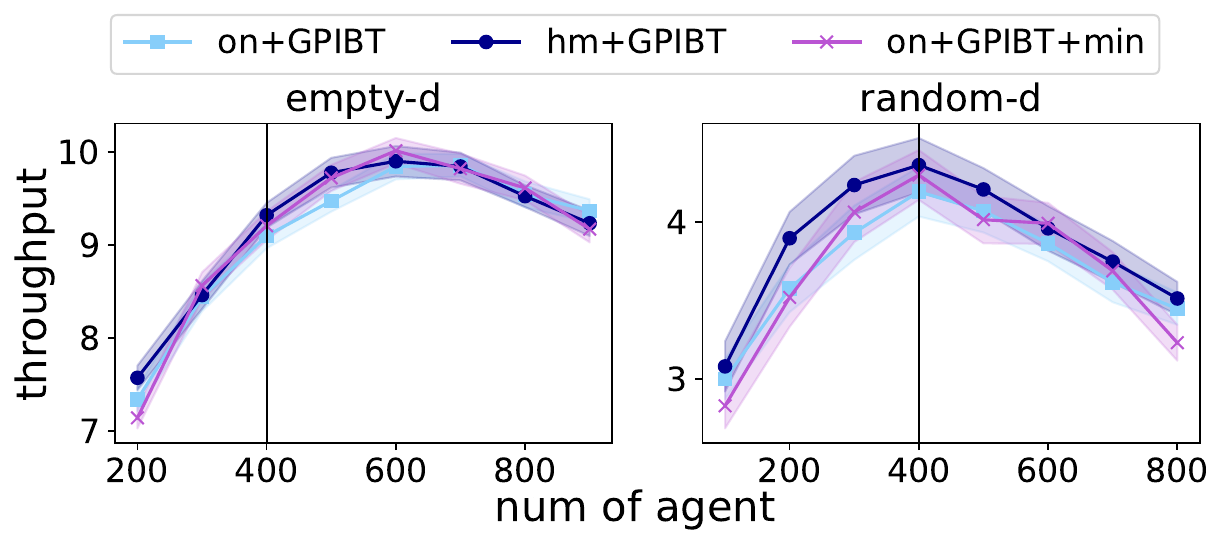}
    \caption{\textit{throughput-agent} curves for hm+GPIBT, on+GPIBT, on+GPIBT+min. }
    \label{fig:min}
\end{figure}

We conducted experiments on the \textit{empty} and the \textit{random} map. Since the underlying graphs for these maps are grid-based, each vertex has at most 4 neighbors. Consequently, the number of parameters in the guidance policy is reduced to 48, far fewer than the original 560. Despite this constraint, the performance of on+GPIBT (denoted as on+GPIBT+min) remains comparable to that of hm+GPIBT. Based on these results, we conjecture that, in these settings, hm+GPIBT is near optimal. \Cref{fig:min} shows the \textit{throughput-agent} curve. The CMAES hyperparameters for on+GPIBT+min are the same as those used for on+GPIBT.

\label{appendix:result:update}
\begin{figure}[h]
    \centering
    \includegraphics[width=\linewidth]{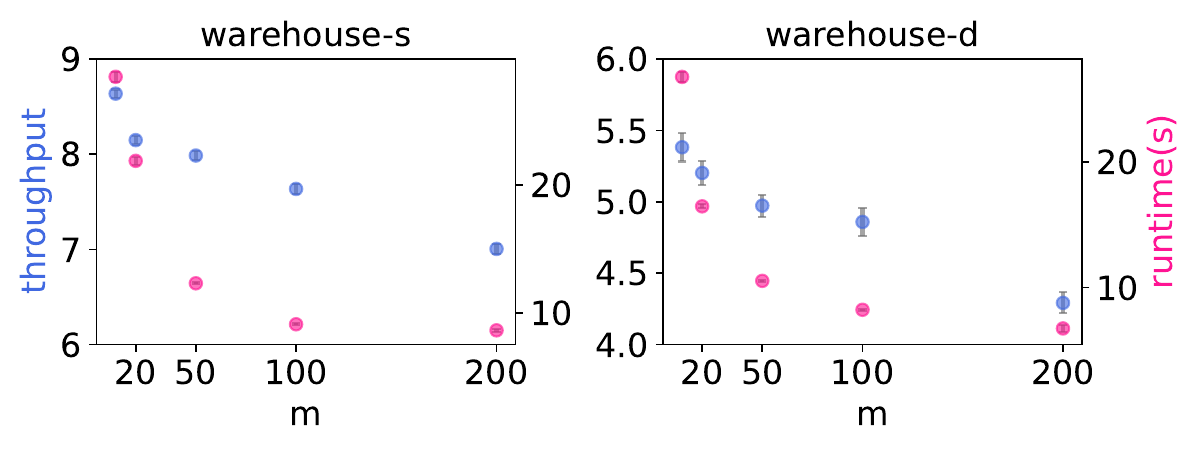}
    \caption{The relationships between throughput and $m$, and the total runtime for 1000 timesteps and $m$, are shown for 600 agents. ``s'' and ``d'' indicate static and dynamic task distribution, respectively. Blue points represent throughput, while pink points represent runtime. Gray bars indicate the 95\% confidence interval, although some of them are too small to be visible. }
    \label{fig:ab-m}
\end{figure}
\subsection{Ablation on Guidance Update Interval} 
\Cref{sec:application} mentioned the guidance update interval $m$ for PIBT. The selection of $m$ is a trade-off between the adaptability of the guidance policy to varying traffic patterns and the computational overhead. Updating the guidance graph more frequently could lead to better throughput at the expense of more computational overhead.

\mysubsubsection{Experiments setup. }
In this section, we compare the performance of $m\in\{10, 20, 50, 100, 200\}$. We optimize all guidance policies with CMA-ES with 600 agents. We focus on the \textit{warehouse} map with static and dynamic task distribution. We set CMA-ES hyperparameters $N_{eval}=25,000$, $b=50$, and $N_{e}=3$.

\mysubsubsection{Results. }
\Cref{fig:ab-m} shows the ablation results. We observe that the smaller the $m$, the higher the throughput and the longer the runtime. Given the trade-off of runtime and throughput with the choice of $m$, it depends on the exact planning time limit of the downstream application scenario of our algorithm to determine the exact value of $m$.

    

\begin{figure*}
    \centering
    \subfigure[\textit{random}, 400 agents]{
    \includegraphics[width=0.15\linewidth]{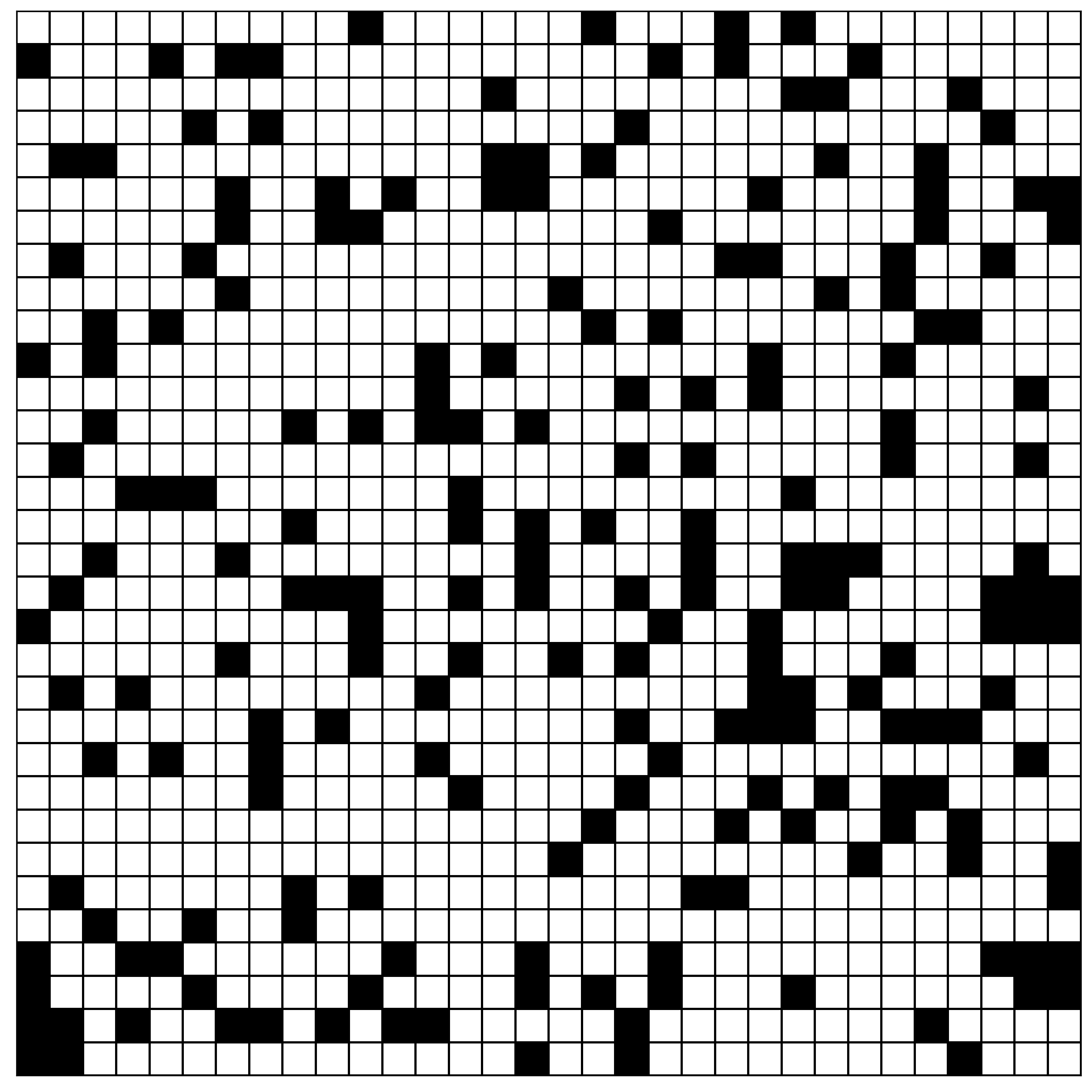}
    
    \includegraphics[width=0.25\linewidth]{
    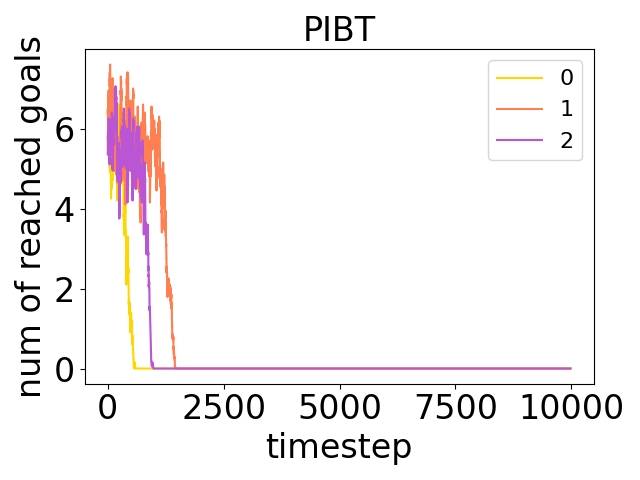
    }
    \includegraphics[width=0.25\linewidth]{
    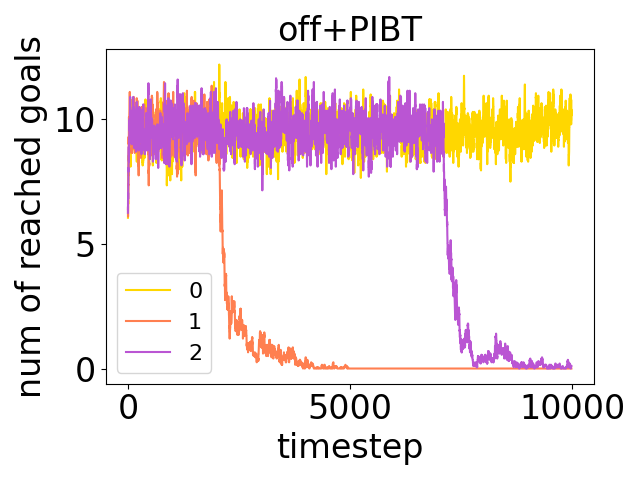
    }
    \includegraphics[width=0.25\linewidth]{
    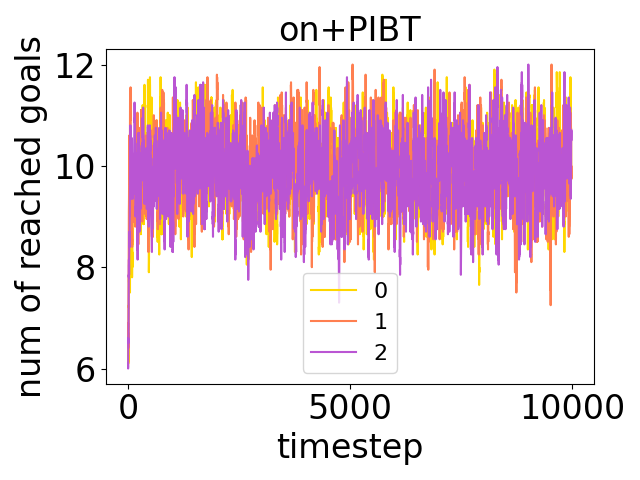
    }
    
    }
    \subfigure[\textit{maze}, 400 agents]{
    \includegraphics[width=0.15\linewidth]{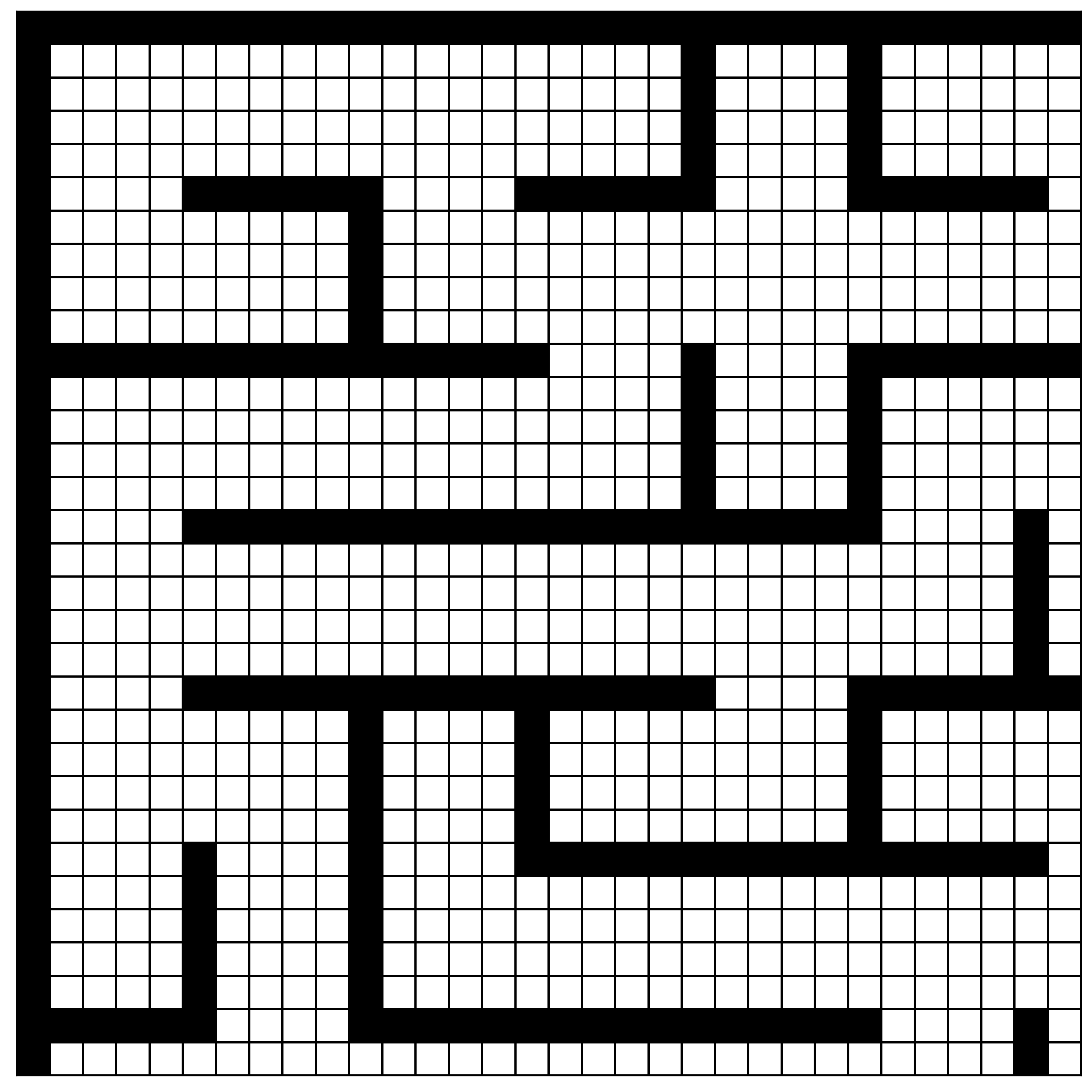}
    \includegraphics[width=0.25\linewidth]{
    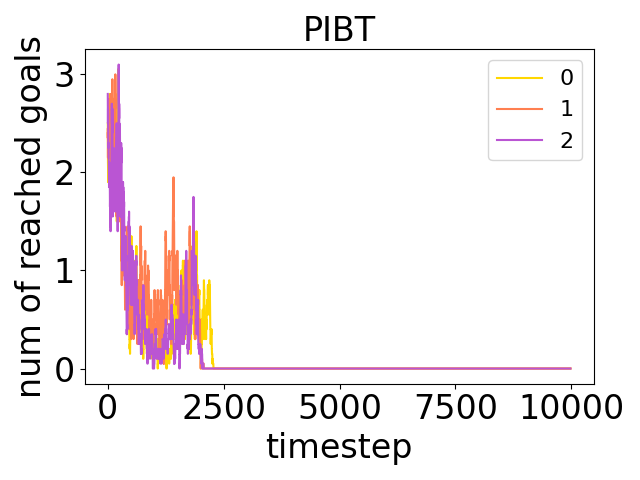
    }
    \includegraphics[width=0.25\linewidth]{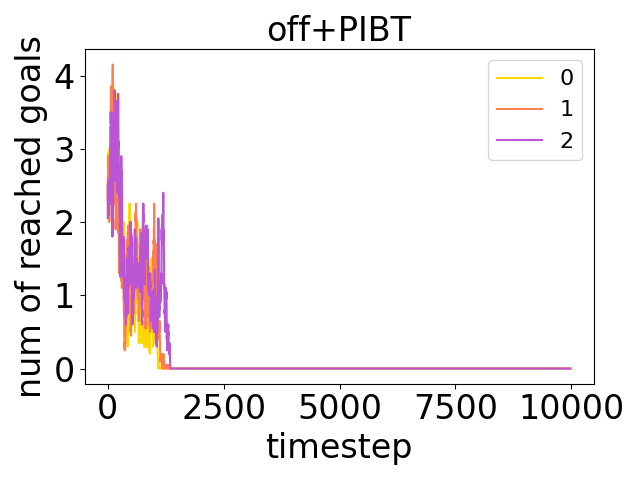}

     \includegraphics[width=0.25\linewidth]{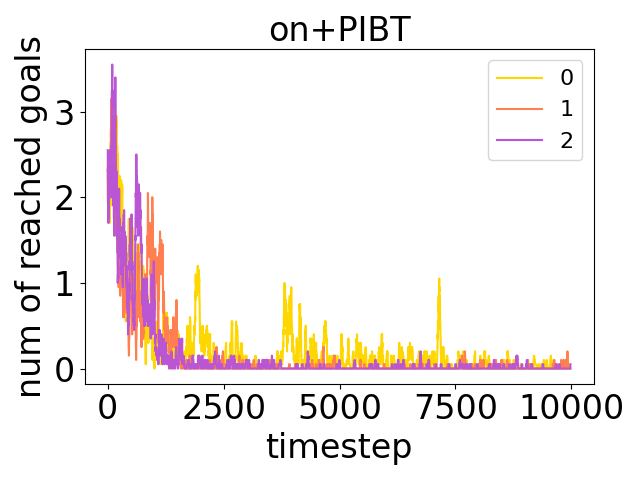}

    }
    \subfigure[\textit{game}, 1600 agents]{
    \includegraphics[width=0.15\linewidth]{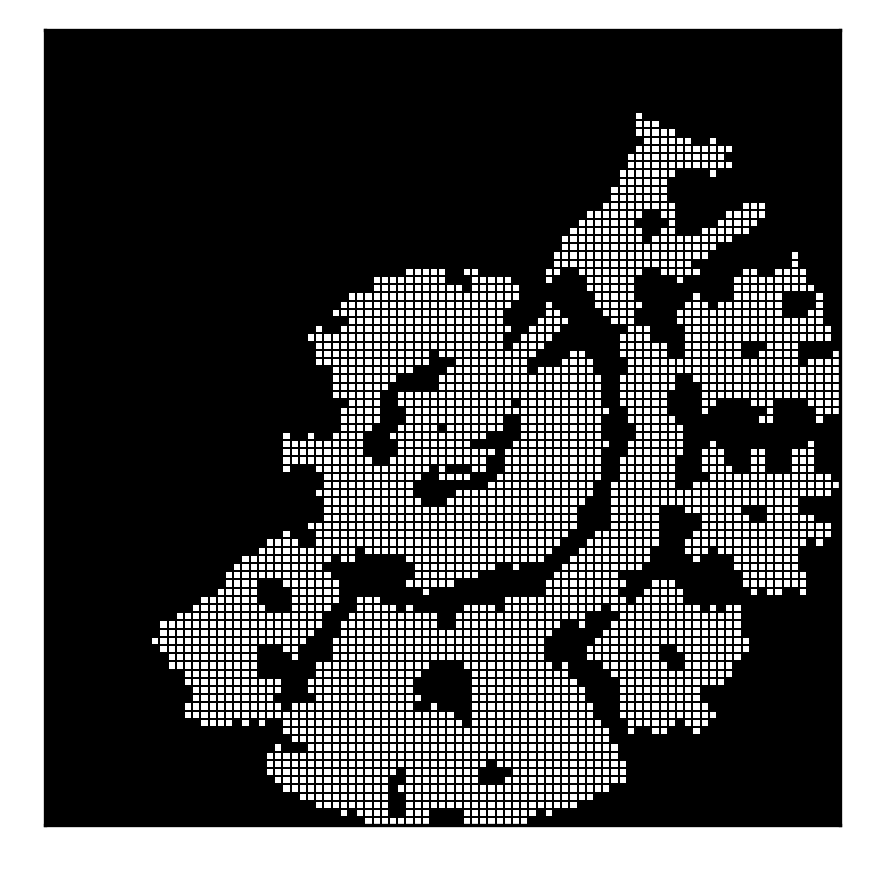}
    \includegraphics[width=0.25\linewidth]{
    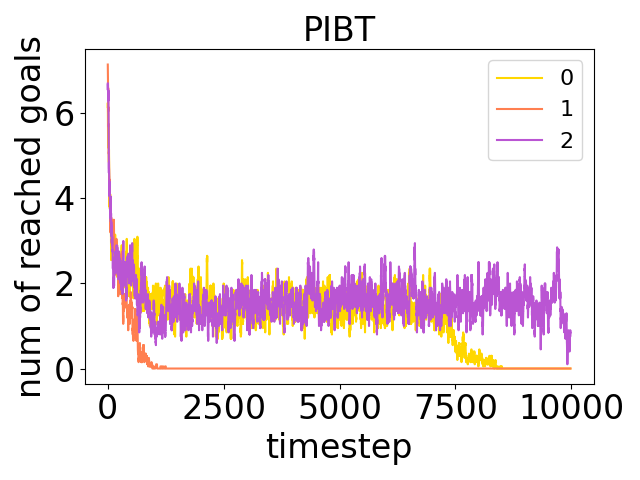
    }
    \includegraphics[width=0.25\linewidth]{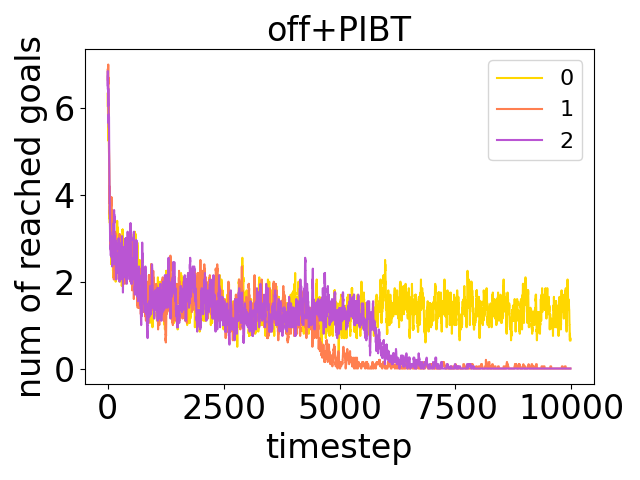}
    \includegraphics[width=0.25\linewidth]{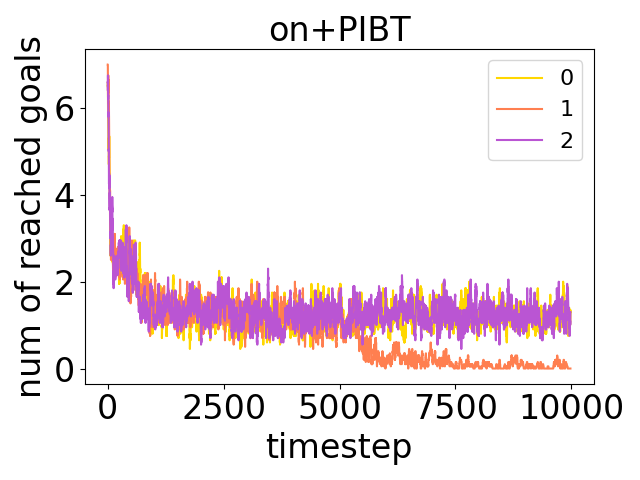}
    }
    \caption{Finished tasks at each step of PIBT with the ``swap'' technique. Curves are smoothed by averaging over 20 timesteps. 0, 1, 2 indicate the seeds of experiments.}
    \label{fig:deadlock-2}
\end{figure*}
\subsection{Deadlock Issues}
\label{appendix:deadlock}
Despite using the ``swap'' technique, PIBT cannot fully resolve the deadlock issue; it only reduces the likelihood of occurrence. However, in lifelong problems where algorithms are expected to run continuously, even a low failure rate in a single instance of MAPF can accumulate over time. To demonstrate the deadlock issue, we conduct experiments with PIBT over 10,000 timesteps. The visualization results are shown in \Cref{fig:deadlock-2}.

We conduct experiments on the \textit{random-32-32}, \textit{maze-32-32}, and \textit{game-small-97-97} maps. The \textit{random} map is the same as the one used in the main results in \Cref{fig:all}, the \textit{maze} map is adopted from the MAPF benchmark~\cite{SternSoCS19}, and the \textit{game-small} map is downsampled from the \textit{game} map shown in \Cref{fig:largemap-result} with modifications to ensure the map is connected.

For the \textit{random} map, it is apparent that effective guidance can help alleviate the deadlock issue. Offline guidance (off+PIBT) can reduce deadlock occurrences to some extent, and our online guidance policy (on+GPIBT) further alleviates the problem. However, the alleviation is limited. For the \textit{maze} map and the \textit{game} map, we do not observe significant improvements.

\Cref{tab:hyperparams,tab:hyperparams-dead} shows the CMA-ES parameters for on+PIBT and off+PIBT. 
\begin{table}[h]
\centering
\resizebox{0.9\linewidth}{!}{
\begin{tabular}{c|c|rrr}
\toprule
map-[dist type] & algorithm           & $N_{eval}$ & $b$  & $N_e$ \\\midrule
\textit{maze}-s  & on+PIBT & 12,500  & 50    & 3 \\
                    & off+PIBT & 12,500 & 50    & 3 \\\midrule
\textit{game-small}-s & on+PIBT & 10,000 & 50  &   3\\
                & off+PIBT & 10,000    & 50  & 3\\
\bottomrule
\end{tabular}
}
\caption{CMA-ES hyperparameters for the \textit{maze} and \textit{game-small} map.}
\label{tab:hyperparams-dead}
\end{table}

\end{document}